\documentclass[twocolumn,amsmath,amssymb,pra,showpacs]{revtex4-2}
\usepackage{graphicx,amsmath,amssymb,amsfonts,latexsym,dcolumn,bm,subfigure,float}
\usepackage{multirow}
\usepackage[usenames,dvipsnames]{color}
\usepackage{siunitx,braket,physics}

\renewcommand{\imath}{\mathrm{i}}
\newcommand{\be}[0]{\begin{equation}}
\newcommand{\ee}[0]{\end{equation}}

\begin{document}

\title{Using graphene conductors to enhance the functionality of atom-chips}

\author{K. Wongcharoenbhorn, R. Crawford, N. Welch, F. Wang, \\ G. Sinuco-Le\'on$^1$, P. Kr\"uger$^2$, F. Intravaia$^3$, C. Koller$^4$, T.M. Fromhold}

\affiliation{School of Physics and Astronomy, University of Nottingham, Nottingham NG7 2RD, UK}
\affiliation{$^1$Department of Chemistry, Durham University, South Road, Durham DH1 3LE, UK}
\affiliation{$^2$Department of Physics and Astronomy, University of Sussex, Brighton BN1 9QH, UK}
\affiliation{$^3$ Humboldt-Universit\"at zu Berlin, Institut f\"ur Physik, 12489 Berlin, Germany}
\affiliation{$^4$University of Applied Sciences Wiener Neustadt, Johannes Gutenberg-Stra{\ss}e 3, 2700 Wiener Neustadt, Austria}

\date{\today}

\begin{abstract}
We show that the performance and functionality of atom-chips can be transformed by using graphene-based van der Waals heterostructures to overcome present limitations on the lifetime of the trapped atom cloud and on its proximity to the chip surface. Our analysis involves Green-function calculations of the thermal (Johnson) noise and Casimir-Polder atom-surface attraction produced by the atom-chip. This enables us to determine the lifetime limitations produced by spin-flip, tunneling and three-body collisional losses. Compared with atom-chips that use thick metallic conductors and substrates, atom-chip structures based on two-dimensional materials reduce the minimum attainable atom-surface separation to a few 100 nm and increase the lifetimes of the trapped atom clouds by orders of magnitude so that they are limited only by the quality of the background vacuum. We predict that atom-chips with two-dimensional conductors will also reduce spatial fluctuations in the trapping potential originating from imperfections in the conductor patterns. These advantages will enhance the performance of atom-chips for quantum sensing applications and for fundamental studies of complex quantum systems.  
\end{abstract}

\maketitle

\section{Introduction}




Cold atom systems play a key role in both fundamental and applied aspects of quantum sensing as they provide a well isolated and controllable platform, whilst still being sensitive to fundamentally interesting interactions such as gravity or magnetic fields \cite{Bloch2008,Degen2017}. The high level of uniformity and homogeneity of cold atomic ensembles also provides a platform for high-accuracy time standards \cite{bidel2013,bize2005}. Recent experiments have demonstrated atomic quantum sensors in precision accelerometers \cite{Gustavson1997}, clocks \cite{Elgin2019}, and in measuring magnetic fields with an unprecedented combination of high sensitivity (nT), spatial resolution ($\SI{}{\micro\metre}$), and field of view ($\sim \SI{100}{\micro\metre}$) \cite{wildermuth2005,wildermuth2006,Patent,Romalis2007,Lev1,Lev2,Shah2018,Lev3}. As a result, there is now worldwide activity on the development of cold-atom based quantum sensing and timing technologies \cite{UKRev,EURev}. \\ 

Miniaturizing and integrating cold-atom quantum systems for fundamental experiments and technology development has advanced through the creation of atom-chips, which use micro-fabricated current-carrying wires to trap and control the atoms in an ultra-high vacuum and typically $1$-$\SI{100}{\micro\metre}$ from the chip surface. Such chips enable coherent manipulation of the atoms' internal and external degrees of freedom \cite{Folman2002,Fortagh2007} leading, for example, to on-chip formation of Bose-Einstein Condensates (BECs) \cite{ott2001, Hansel2001}, atom interferometers \cite{schumm2005, Wang2005}, and interfacing quantum gases with nanomechanical oscillators \cite{Hunger_Camerer_2010}, carbon nanotubes \cite{schneeweiss2012} and cryogenic surfaces \cite{Huf2009,dikovsky2009,Nirren2006,Roux2008,Cano2011,Minniberger2014}. However, commonly-used metal wires with a typical thickness of $\sim \SI{1}{\micro\metre}$, mounted on bulk insulating substrates, have adverse effects when trapped atom clouds approach the surface. Spatial imperfections in the wires roughen the trapping potential, Johnson noise currents induce spin-flip transitions that eject atoms from the trap, and the strong Casimir-Polder (CP) attraction between the atoms and the chip produce tunneling losses. Together, these loss mechanisms prevent the formation of long-lived microtraps at distances closer than several microns from the chip surface \cite{Henkel2003, Lin2004}.\\

Overcoming these limitations is needed to advance both the fundamental and technological applications of micro- and nano-engineered environments for cold atoms. Trapping atoms closer to the chip offers a number of advantages. Higher magnetic field gradients and trap frequencies can be attained for a given current, thereby facilitating fast initial cooling, i.e. before three-body collisions become relevant, as required for creating BECs under less stringent vacuum requirements. Higher trapping frequencies will also produce atomic gases that are closer to the 1D limit and thus better for studying the thermodynamics of low-dimensional gases.  Sub-micron trapping was realized by utilizing the balancing of attractive and repelling forces of light in nano-fibres \cite{Vetsch2010} \cite{schneeweiss2014}, but has proven difficult for magnetic trapping \cite{Meng_2018}. Enabling this offers a pathway to creating hybrid quantum devices comprising coherently-coupled atomic and solid-state elements \cite{Verdu2009,Bernon2013}.\\ 

Achieving long lifetimes for atom clouds trapped within $\SI{1}{\micro\metre}$ of the chip surface will enable quantum gases to be controlled, entangled, and addressed by potential landscapes whose spatial features are finer than the intrinsic length scales of atomic gases, for example the healing length. Sub-micron atom-surface trapping distances will also advance chip-based sensors in which atoms are used to measure external fields and forces. One such sensor, the BEC microscope, can image current flow patterns in planar conductors with a spatial resolution limited primarily by the distance of the BEC from the conductor \cite{wildermuth2005,wildermuth2006,Patent,Lev1,Lev2,Lev3}.\\

Here, we show that atom-chips containing two-dimensional conductors can, in principle, overcome the present limitations on the atom-surface separation and lifetime of the trapped atom cloud. Such trapping structures can be fabricated using, for example, graphene membranes that are either free standing or enclosed by two-dimensional insulating layers so as to form a van der Waals heterostructure \cite{vdW1,vdW2}. This opens a route to achieving sub-micron trapping distances and, hence, fine features in the trapping potential landscape that are not attainable when conventional metallic wires are used. We demonstrate that van der Waals heterostructures can be used to form traps just a few hundred nanometers from their surface whilst maintaining trap lifetimes of at least ten seconds. This exceeds the duration of most experiments on the atom clouds and of typical active operation cycles in cold-atom quantum sensors. In previous work on the possible use of two-dimensional electron gases as conductors in atom chips \cite{SinucoA,SinucoB}, the lifetime of nearby atomic gases was estimated by extrapolating from the rates of tunneling losses and Johnson noise-induced spin flips near metallic conductors \cite{SinucoB}. Here, we present detailed calculations of the atom cloud lifetimes, in which the same Green function formalism is used to determine both the CP potential and Johnson noise lifetimes, thereby ensuring a fully consistent picture of atom loss rates.\\ 

The paper is organized as follows: In Section II, we describe the layers of two-dimensional materials that comprise the atom-chip and explain how the chip affects the lifetime and minimum practical atom-surface separation of trapped atom clouds. In Section III, we compare the limiting factors, specifically atom cloud lifetime and spatial roughness, for traps formed less than $\SI{1}{\micro\metre}$ away from the surface of chips containing graphene or metallic trapping wires. Specifically, we present detailed calculations that quantify how two-dimensional conductors such as graphene can reduce both the spin-flip atom losses resulting from Johnson noise in the conductor and the tunneling losses due to CP atom-surface attraction sufficiently to enable stable sub-micron trapping with atom-cloud lifetimes $> 10$ s. In Section IV, we propose specific routes to the realization of graphene-based atom-chips that operate under realistic experimental conditions. Finally, in Section V we conclude with an overview of possible further device geometries and experiments to demonstrate the performance and versatility of graphene-based atom-chips.\\

\section{atom-chips: structure and influence on trapped atoms}\label{sec:lifetime}

\begin{figure}[ht]
\includegraphics[width = \linewidth]{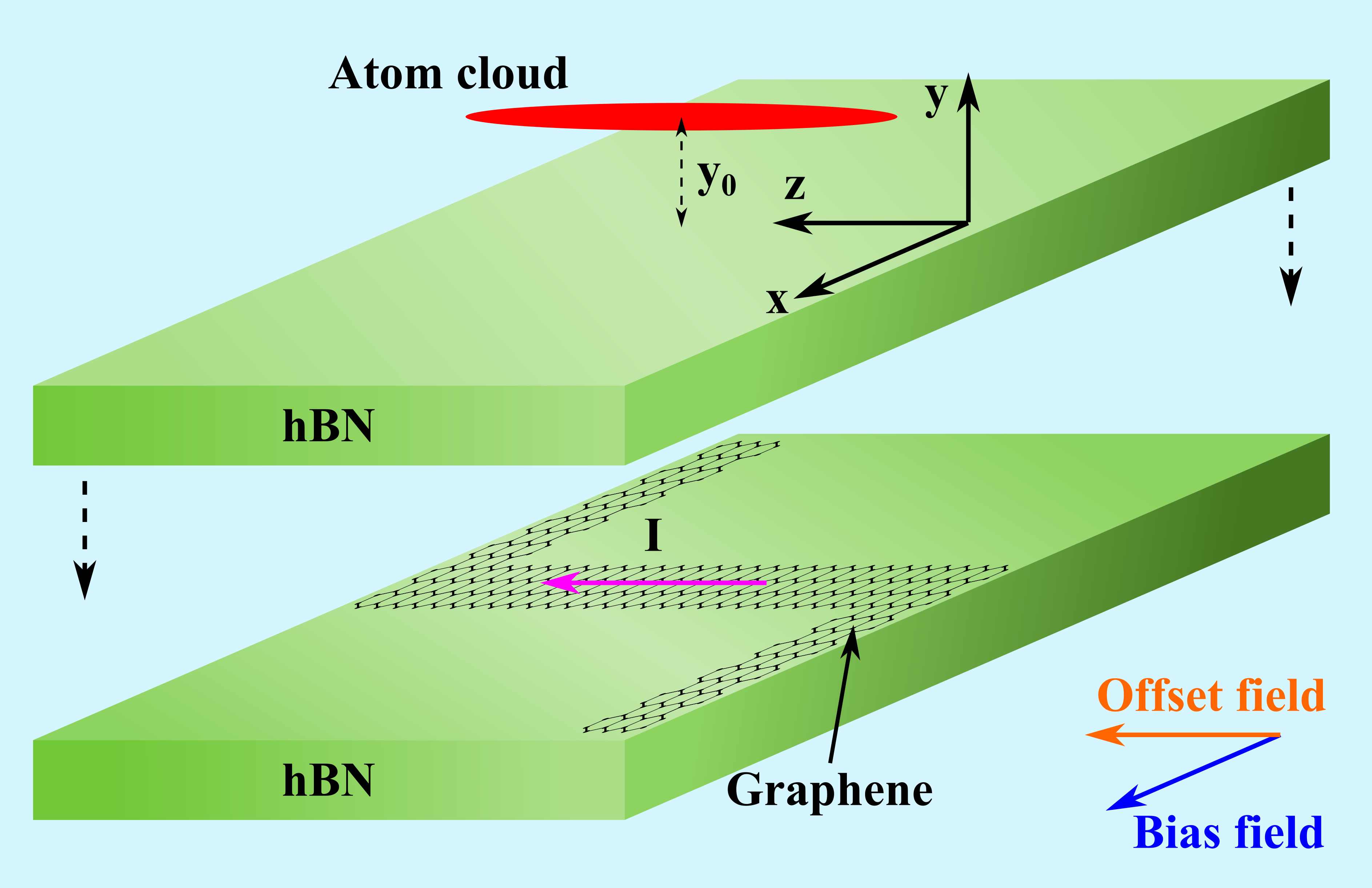}
\caption{Schematic diagram of the proposed graphene-based atom-chip showing the Z-shaped graphene conducting channel (hexagonal graphene lattice pattern) carrying current, $I$ (pink arrow), encased by thin, protective, hBN cladding layers (upper and lower green slabs). The orientations of the applied magnetic bias field and the offset field are shown by blue and orange arrows respectively. These fields combine with that produced by the conducting channel (current $I$) to trap a nearby atomic BEC (red).}
\label{fig:chip_diagram}
\end{figure}

We consider micro-fabricated atom-chip structures that produce magnetic traps for clouds of rubidium-87 ($^{87}$Rb) atoms, as this approach will facilitate comparison to previous experiments. The implications for other species of alkali atoms are straightforward to derive and do not differ qualitatively from the $^{87}$Rb case. Our proposed graphene-based atom-chip is shown schematically in Fig.~\ref{fig:chip_diagram}. The chip comprises a Z-shaped graphene wire encased by two cladding layers of $\SI{10}{\nano\metre}$-thick hexagonal boron nitride (hBN). Electrical current, $I$, through the Z-shaped wire generates an inhomogeneous magnetic field, which is supplemented by a constant applied bias field, $\mathbf{B}_{b}$, and an offset (Ioffe) field, $\mathbf{B}_{0}$, to create a magnetic field minimum at $\mathbf{r}_{0} = (x_{0}, y_{0}, z_{0})$. An ultracold atom cloud is trapped near this magnetic field minimum, whose value is non-zero due to the offset field, which suppresses atom losses due to Majorana spin-flip transitions \cite{Folman2002}. The potential energy, $U_{\mathrm{mag}}$, of the trapped atoms equals the interaction energy between the atomic magnetic moment $\boldsymbol{\mu}$ and the net magnetic field, $\mathbf{B}(\mathbf{r})$, where $\mathbf{r}$ is the spatial position with respect to the coordinate origin, i.e.
\begin{equation}
U_\mathrm{mag}(\mathbf{r}) = -\boldsymbol{\mu}\cdot\mathbf{B}(\mathbf{r}) = m_{F} \mu_{B} g_{F} \abs{\mathbf{B}(\mathbf{r})}.
\label{equ:MagPot}
\end{equation}
Here, $\mu_{B}$ is the Bohr magneton and $g_F$ is the Land\'e factor of the relevant hyperfine state. For the $^{87}$Rb atoms considered here, this is typically the $\ket{F, m_{F}} = \ket{2, 2}$ level of the $5^2S_{1/2}$ ground state. Provided that the magnetic quantum number, $m_F$, is a good quantum number, atoms in metastable low-field seeking states, whose magnetic moment is aligned anti-parallel to the magnetic field orientation, will be trapped near the magnetic field minimum, $\mathbf{r}_{0}$, where $U_{\mathrm{mag}}$ is also minimal.\\

The interplay between three energy scales determines the lifetime of the trapped atomic gas. The first energy scale is the trap depth given by $V_{0} = \abs{\boldsymbol{\mu}\cdot\Delta\mathbf{B}}$, where $\Delta\mathbf{B}$ is the difference between the maximum and the minimum values of the magnetic field. The second scale is the thermal energy, $k_{B}T_{\mathrm{cloud}}$, related to the temperature, $T_{\mathrm{cloud}}$, of the trapped atoms, where $k_{B}$ is the Boltzmann constant. The last energy scale is the ground-state energy of the trapped atoms, $E_{0}$. Provided that a sufficiently deep magnetic trap, i.e. $ V_{0} \gg k_{B}T_{\mathrm{cloud}}$, $E_{0}$, is located far away from any surface the overall lifetime of the trapped atoms is limited by collisions with the background gas. In typical atom-chip experiments, pressures of $10^{-10}$ to $10^{-11}$ mbar or below can be reached, for which the background pressure-limited lifetime of the trapped atoms is of the order of tens of seconds or better \cite{Folman2002}. As the atom cloud approaches the chip surface, its lifetime is reduced by modification of the trapping potential due to CP interactions with the surface, see Figs. \ref{fig:dipole_interface} and \ref{fig:CP_Gr_hBN_Gold},  and Johnson noise in the conductor, which can cause the atoms to undergo spin-flip transitions into untrapped states. Lower-frequency Johnson noise, comparable with the trap frequencies, can also potentially cause atom losses due to parametric heating of the atom cloud. However, the rates of such heating are orders of magnitude lower than the spin-flip loss rates \cite{Henkel2003}. Due to the low Johnson noise and CP potential near graphene-based atom-chips, the lifetime of atom clouds trapped near such chips will, beyond a certain trapping distance, be limited only by the background pressure as we quantify below.\\ 

As the atom-surface trapping distance decreases, the trap frequencies must be increased in order to reduce depletion by the CP interaction. In turn, this increases the density of the trapped atom cloud and, therefore, also increases the rate of three-body collision losses discussed in Section III. In order to determine the optimal trapping distance, the interplay of three-body losses, the minimum detection density of the atom cloud, and the CP interaction all have to be considered, as discussed in Section III. 

\subsection{CP potential and resulting atom tunneling towards the chip surface}\label{sec:tunnelling towards the chip surface}

The Casimir-Polder potential is essentially a position-dependent shift of the atomic energy level structure, induced by the interaction of the atom with the surrounding surface-modified electromagnetic radiation \cite{casimir1948,Wylie84}. In general, the presence of an object modifies a system's electromagnetic density of states, due to the boundary conditions that the field has to satisfy on the surface of the object. The extent of the modification, and therefore the strength of the CP potential, depends on the object's specific position in space, on its form and on the material(s) from which it is made.\\

Generally, an atom experiences an attractive CP force towards metallic and dielectric surfaces. In atom-chip systems this behavior effectively lowers the barrier at the side of a magnetic trap that is nearest the surface, as shown in Fig. \ref{fig:CP_many_structures}. In turn, this enables atoms to tunnel out of the trap and be lost from the atom cloud. tunneling losses induced by the CP potential affect key atom-chip performance parameters such as the integration time for sensor applications and the coherence time for quantum memories. In the present generation of atom-chips, metal wires used to generate the magnetic field 
lead to a large CP attraction on trapped atoms located within $\approx \SI{1}{\micro\metre}$ of the surface. This imposes a minimum trapping distance of $10$-$\SI{100}{\micro\metre}$ for typical atom-chip experiments \cite{Folman2002, Jones2003, Harber2003}. 
Our proposed 2D material-based atom-chips are expected to exert very low CP attraction, due to their extremely small ($<$ 100 nm) thickness and their specific material properties, thereby opening a new route to entering the sub-micron atom-surface trapping regime.  

\begin{figure}[ht]
\centering
\includegraphics[width = \linewidth]{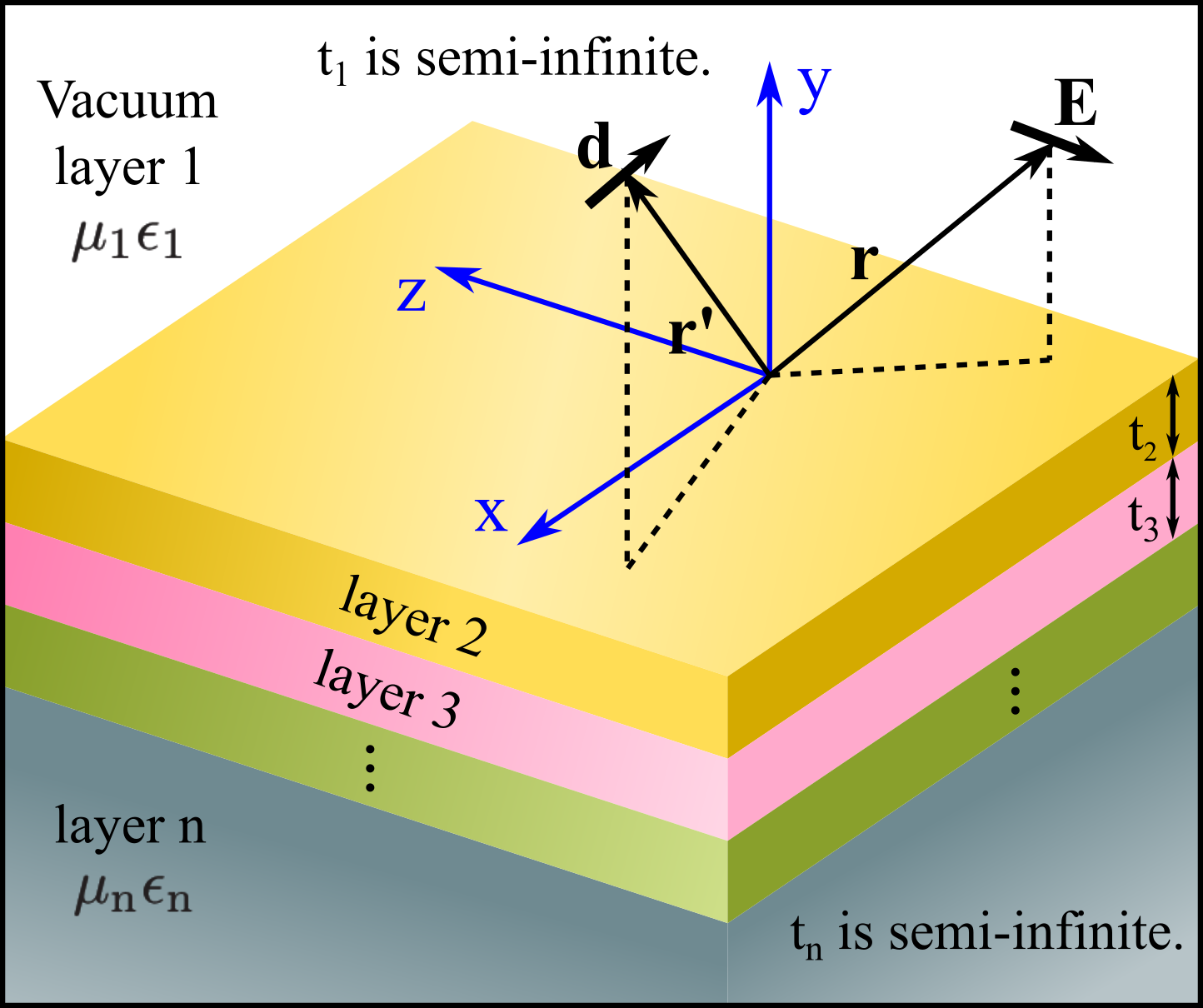}
\caption{Schematic diagram of a dipole (electric in the case of our CP potential calculation but magnetic for our Johnson noise analysis) near an $n$-layer system, where each layer is designated by index $l = 1, 2,..., n$. Each layer is characterized by thickness, $t_{l}$, permeability, $\mu_{l}$, and permittivity, $\epsilon_{l}$. The top surface of upper material layer 2 (yellow) coincides with the origin of the $y$ coordinate. The dipole (depicted by the arrow labeled $\mathbf{d}$), is located at $\mathbf{r}^{\prime} = (x^{\prime}, y^{\prime}, z^{\prime})$ in layer 1, above the solid material layers, and acts as a point source. The arrow labeled $\mathbf{E}$, represents the orientation of the electric field of frequency, $\omega$, at point $\mathbf{r}$, which is related to the dipole $\mathbf{d}$ via a total Green's tensor. Note also that $t_{1}$ and $t_{n}$ are infinite, corresponding to semi-infinite top and bottom layers.}
\label{fig:dipole_interface}
\end{figure}

For a system in equilibrium at a temperature $T$ consisting of an atom located at position $\mathbf{r}^{\prime}$ from a nearby material body (see Fig.~\ref{fig:dipole_interface}), both interacting with the electromagnetic field,  
the CP potential is given by \cite{Buhmann_thermal,Intravaia11}
\begin{equation}
U_{CP}(\mathbf{r}^{\prime}) = \mu_{0}k_{B}T\sum_{j=0}^{\infty}{}^{'}\xi_{j}^{2}\alpha(\mathrm{i}\xi_{j})\,\mathrm{tr}\big[\mathbf{G}^{(1)}(\mathbf{r}^{\prime}, \mathbf{r}^{\prime}, \mathrm{i}\xi_{j})\big],
\label{eq:CP_for_flatsurface}
\end{equation}
where $\mu_{0}$ is the permeability of vacuum, $\hbar$ is the reduced Planck constant $\alpha(\omega)$ is the atomic polarizability and $\xi_{j} = 2\pi k_{B}Tj/\hbar$, are commonly known as the Matsubara frequencies \cite{Matsubara1955}. The prime on the Matsubara sum in Eq.~\eqref{eq:CP_for_flatsurface} indicates that the $j = 0$ term carries half weight \cite{Intravaia11}.
In Eq.~\eqref{eq:CP_for_flatsurface}, $\mathbf{G}^{(1)}(\mathbf{r}^{\prime}, \mathbf{r}^{\prime}, \omega)$ is the scattering Green's tensor, which contains the information about the material's optical properties and the geometry of the system.\\

For atom-surface separations shorter than the size of the components of an actual atom-chip (typically of the order of a few tens of micrometers or larger) we can consider that the atoms are interacting with a large layered surface. In this case the expression for the Green tensor is known and we present it explicitly in Appendix \ref{supplement: Green's function}. As shown there, in order to determine the expression of the Green tensor we need to evaluate the reflection coefficients of the electromagnetic field incident on the atom-chip structure. In our case, the multi-layer configurations allow their determination using the scattering or the transfer-matrix approach \cite{Yariv83,Zhan2013} in combination with models describing the optical properties of graphene, hBN and gold (see Appendix \ref{supplementary: Optical properties} for details). In this work, we take the Fermi energy and electron relaxation rate of graphene to be $E_{F} = \SI{0.1}{eV}$ and $\gamma = \SI{4}{THz}$, respectively, corresponding to typical values found both theoretically \cite{gric_2019,andryieuski_lavrinenko_2013,Goncalves2016,amorim_2017} and in experiments \cite{ju_geng_horng_2011}.\\

In this paper, we consider a simple model of alkali atoms with atomic polarisability of the form \cite{JuddA,Intravaia11}
\begin{equation}
\alpha(\mathrm{i}\xi_{j}) = \alpha_{0}\frac{\omega_{\mathrm{T}}^{2}}{\omega_{\mathrm{T}}^{2} + \xi_{j}^{2}},
\label{eq:polarisability}
\end{equation}
where $\alpha_{0}$ is the ground-state static polarizability and $\omega_{\mathrm{T}}$ is the frequency of the dominant atomic transition. In the case of \textsuperscript{87}Rb atoms, $\alpha_{0} = \SI{5.27e-39}{\farad\metre^{2}}$ \cite{Schwerdt2018}, and $\omega_{\mathrm{T}} = 2\pi\times\SI{384}{THz}$ is the D2 line transition frequency corresponding to a wavelength of $\SI{780}{\nano\metre}$ \cite{Steck2020}.

Eq. \eqref{eq:CP_for_flatsurface} is sufficiently generic to enable the CP potential to be calculated for our atom-chip system and compared consistently with other materials and structures.\\

\begin{figure}[ht]
\includegraphics[width = \linewidth]{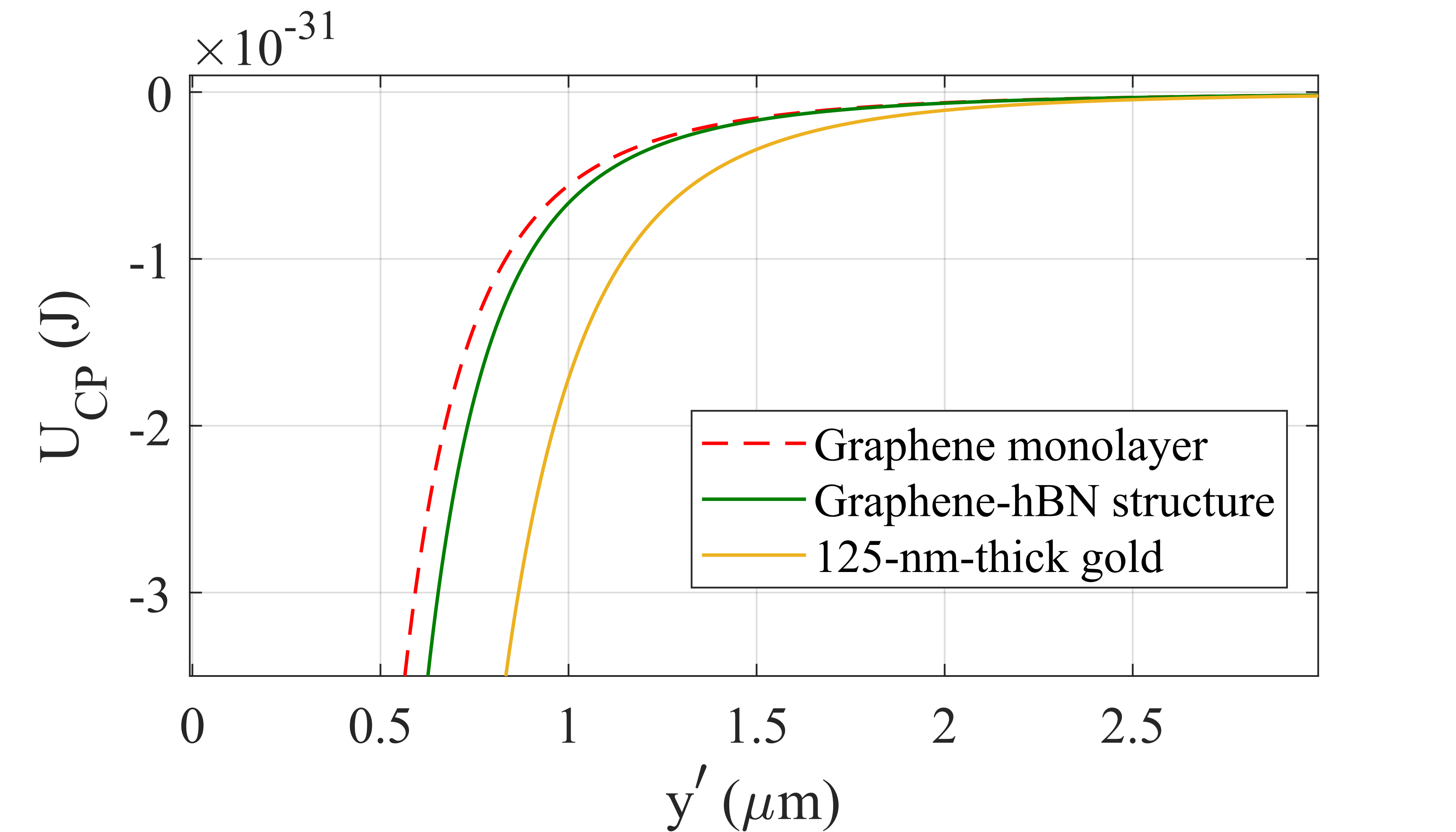}
\caption{CP potential, $U_{CP}$, calculated versus the position, $y^{\prime}$, of an \textsuperscript{87}Rb atom from the surface of: a graphene monolayer (dashed red curve), a heterostructure comprising a graphene monolayer encased by two $10$-$\SI{}{\nano\metre}$-thick hBN layers (solid green curve), a $125$-$\SI{}{\nano\metre}$-thick gold slab (solid yellow curve) at temperature $T = \SI{300}{\kelvin}$.}
\label{fig:CP_Gr_hBN_Gold}
\end{figure}

In Fig.~\ref{fig:CP_Gr_hBN_Gold}, we compare the CP potential, $U_{CP}$, calculated versus the separation, $y^{\prime}$, of an \textsuperscript{87}Rb atom from three different material systems: a graphene monolayer (dashed red curve) for which CP potential calculations have been reported previously \cite{Churkin,MosteA,MosteB,JuddA,ScheelA,AntezzaA,mostepanenko_2020}, a heterostructure comprising a graphene monolayer encased by two $10$-$\SI{}{\nano\metre}$-thick hBN layers (solid green curve) and a $125$-$\SI{}{\nano\metre}$-thick gold slab (solid yellow curve), all at $T = \SI{300}{\kelvin}$. We choose the thickness of the gold slab to be $125$-$\SI{}{\nano\metre}$ because this is among the smallest reported thicknesses at which gold wires in atom-chip experiments \cite{salem_japha2010} have a conductivity that still behaves as bulk. Even in this limit of metallic conductor thickness, at $y^{\prime} \approx \SI{1}{\micro\metre}$, the CP potential for the heterostructure is approximately 40$\%$ of that for the thin gold slab.\\


The effect of the CP potential on the total trapping potential, $U_{\mathrm{tot}}$, can be illustrated by modelling the magnetic trapping potential, $U_H$,  as simple harmonic and adding the CP potential, giving
\begin{equation}
U_{\mathrm{tot}}(y) = U_{\mathrm{H}}(y) + U_{\mathrm{CP}}(y).
\label{eq:U_tot}
\end{equation}
The simple harmonic potential takes the form 
\begin{equation}
U_{\mathrm{H}}(y) = \frac{1}{2}m\omega_{r}^{2}(y-y_{c})^{2},
\label{eq:U_Harmonic}
\end{equation}
where $m$ is the mass of the trapped atom, $\omega_{r}$ is the radial trapping frequency, and $y_{c}$ is the position of the centre of the simple harmonic trap measured from the surface; note that $y_{c}$ is not necessarily equal to the minimum of the total potential.\\

\begin{figure}
\centering
\includegraphics[width =\linewidth]{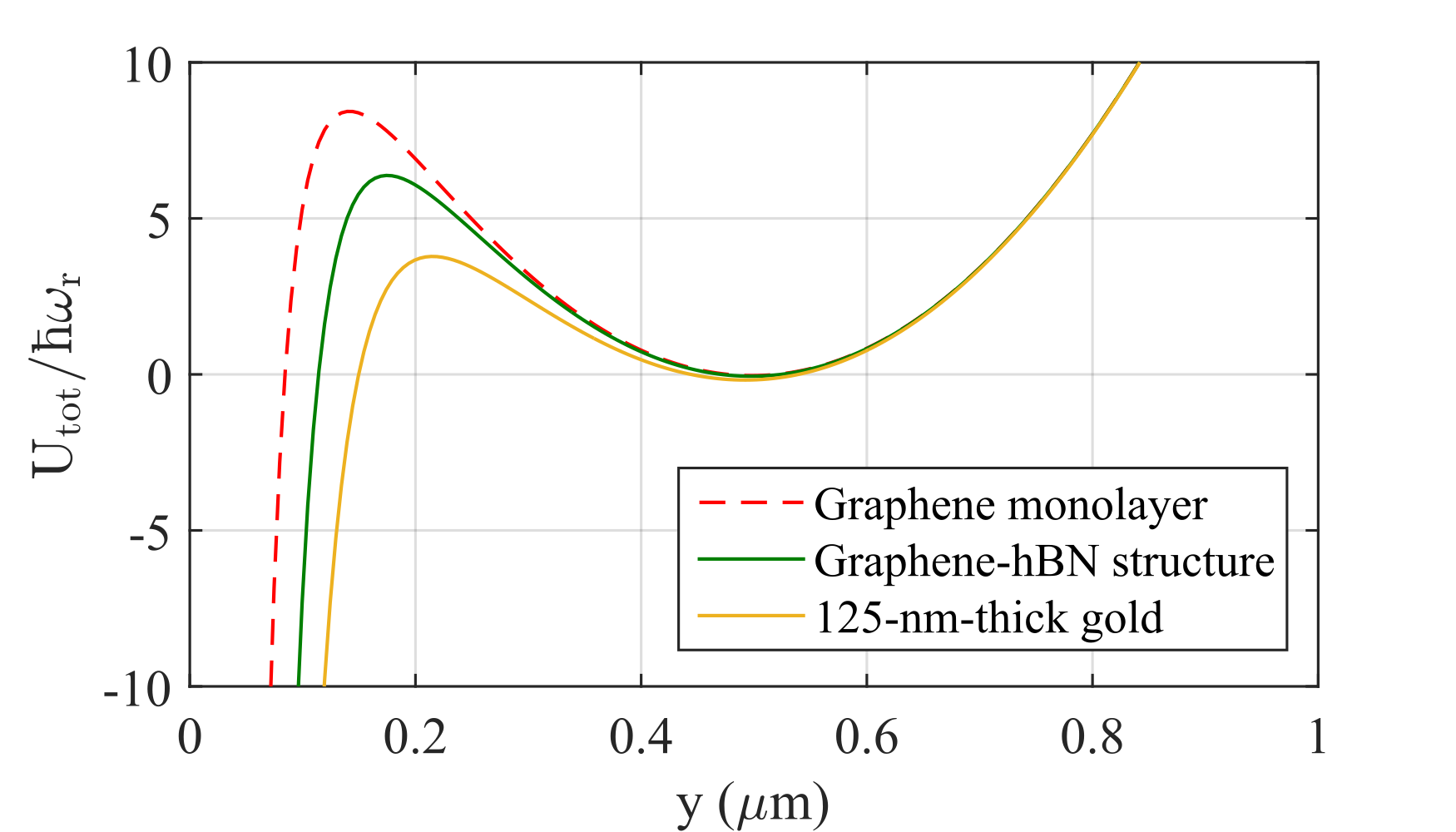}
\caption{Total potential, $U_{\mathrm{tot}}$, calculated versus distance, $y$, of an \textsuperscript{87}Rb atom from the surface of: a free-standing graphene monolayer (dashed red curve); a graphene monolayer encased on each side by a $10$-$\SI{}{\nano\metre}$-thick hBN sheet (solid green curve); a $125$-$\SI{}{\nano\metre}$-thick gold sheet (solid yellow curve). The total potential is the sum of the CP potential and the harmonic model trapping potential, which is centred at $y = \SI{0.5}{\micro\metre}$ with radial trapping frequency, $\omega_{r} = 2\pi\times\SI{20}{kHz}$. All curves are for $T = \SI{300}{\kelvin}$.}
\label{fig:CP_many_structures}
\end{figure}

Fig.~\ref{fig:CP_many_structures} shows the resulting total trapping potential, $U_{\mathrm{tot}}$, calculated versus distance, $y$, of an \textsuperscript{87}Rb atom from a graphene monolayer (dashed red curve), an hBN-graphene monolayer-hBN heterostructure (solid green curve) and a $125$-$\SI{}{\nano\metre}$-thick gold slab (solid yellow curve) calculated taking (see Eq.~\eqref{eq:U_Harmonic}) $\omega_{r} = 2\pi\times\SI{20}{kHz}$, $y_{c} = \SI{0.5}{\micro\metre}$, $T = \SI{300}{\kelvin}$, and the mass of an \textsuperscript{87}Rb atom, $m = \SI{1.44e-25}{kg}$. It is apparent that the CP potential distorts the simple harmonic trap: an energy barrier of finite height and width appears near the surface for $y < y_{c}$. The height and width effectively scale with distance of trap centre from the surface, thereby giving rise to tunneling losses, which deplete the trapped atom cloud. Since graphene creates a weaker CP attraction than even the thin gold conductor, the tunneling loss rates for graphene-based atom-chips are lower than for conventional atom-chips and we quantify this benefit below. Consequently, graphene-based atom-chips offer a performance advantage over the present generation of atom-chips, which use metallic conductors as current-carrying wires.

\subsubsection{Tunneling loss rate}
\begin{figure}
\includegraphics[width = \linewidth]{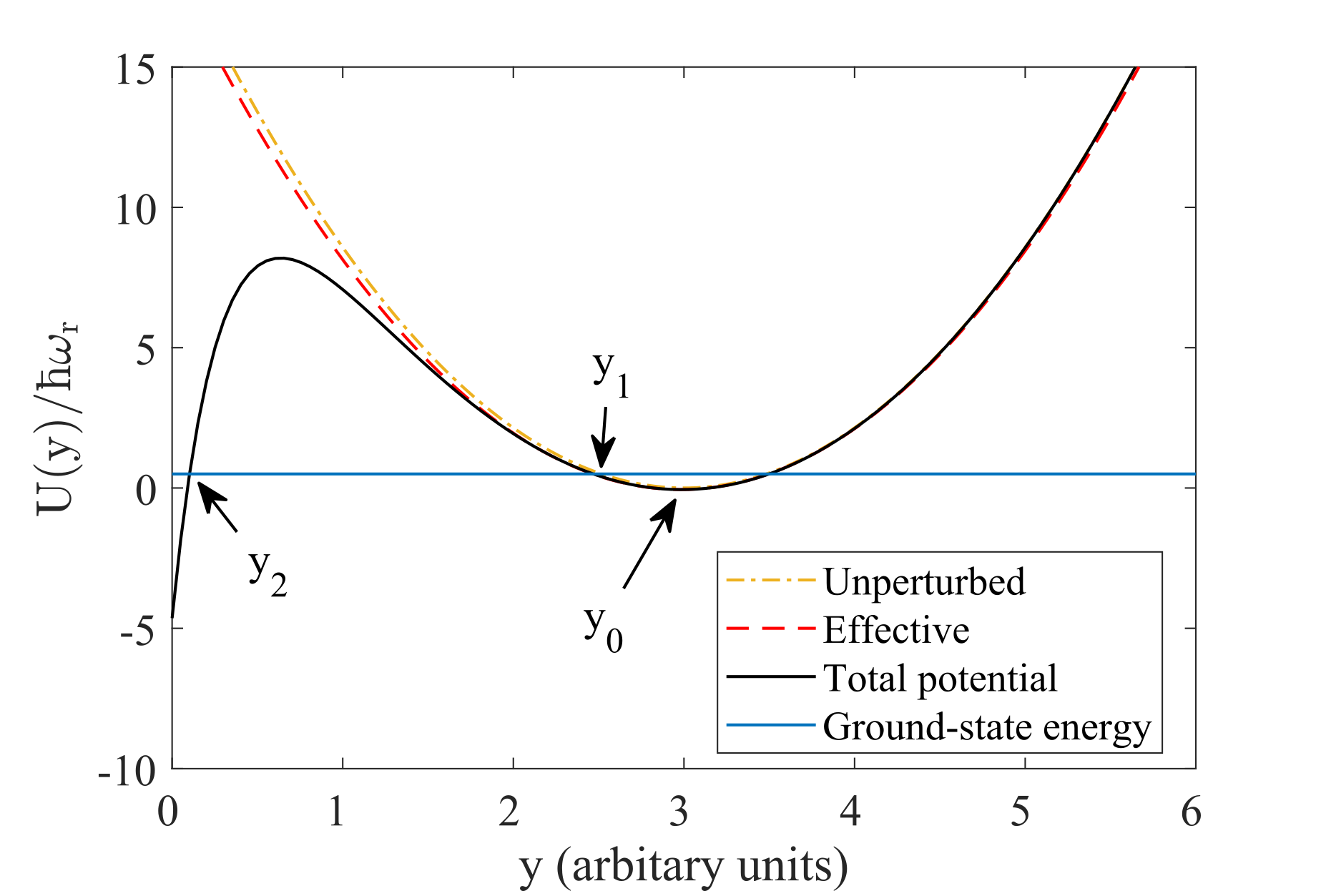}
\caption{Schematic diagram of the total trapping potential, $U_{\mathrm{tot}}(y)$ (solid black curve) plotted versus distance, $y$, of an \textsuperscript{87}Rb atom from an atom-chip surface. The solid blue line indicates the ground-state energy of the quantum harmonic oscillator, $E = \hbar\omega_{\mathrm{eff}}/2$, where $\omega_{\mathrm{eff}}$ is the effective characteristic frequency of the simple harmonic trap (dashed red curve), which is perturbed by the CP potential, as described in the text, and approximates $U_{\mathrm{tot}}(y)$ near the minimum. The positions $y_{0}$, $y_{1}$, $y_{2}$, indicated by arrows, are, respectively, the actual trap centre and the two classical turning points for the left-hand potential energy barrier, where $U(y_{1}) - U(y_{0})  = U(y_{2}) - U(y_{0}) = \hbar\omega_{\mathrm{eff}}/2$. The dash-dotted yellow curve is the potential of the unperturbed simple harmonic trap.}
\label{fig:WKB_2}
\end{figure}

In order to estimate the tunneling loss rate, $\Gamma_{\mathrm{tun}}$, of an atom cloud trapped in the finite potential well shown schematically by the solid black curve in Fig.~\ref{fig:WKB_2}, we employ Gamow's theory of alpha decay \cite{harper_anderson_1997}. In this model, the atom is considered to oscillate inside the potential well and can escape by tunneling through the finite barrier nearest the surface each time it is incident on that barrier. 
The tunneling rate is determined by the frequency at which the atom approaches the barrier, $f$, and the transmission probability $\Tilde{T}$ that the atom tunnels out at each attempt. Mathematically, we have

\begin{equation} 
\Gamma_{\mathrm{tun}} = f\times\Tilde{T}.
\label{eq:tunnelling rate}
\end{equation}

Fig.~\ref{fig:WKB_2} shows that the deformation of the unperturbed harmonic magnetic potential (dot-dashed yellow curve) by the CP interaction also yields an effective perturbed harmonic potential (dashed red curve) with a lower trapping frequency, $\omega_{\rm eff}$, and whose minimum shifts from $y=y_{c}$ to a new position, $y_{0}$. 
The perturbed harmonic potential therefore takes the form
\begin{equation}
U_{\mathrm{eff}}(y) = \frac{1}{2}m\omega_{\mathrm{eff}}^{2}(y-y_{0})^{2},
\label{eq:U_Harmonic_eff}
\end{equation}
where a Taylor expansion of $U_{\rm tot}(y)$ about $y = y_{0}$ yields
\begin{equation}
    y_{0}\approx y_c+\frac{U'_{\rm CP}(y_c)}{m \omega_r^2},
\quad
    \omega_{\rm eff}^2\approx \omega_r^2+U''_{\rm CP}(y_0).
\end{equation}
Atoms in the ground-state of this effective potential have an energy $E = \hbar\omega_{\mathrm{eff}}/2$ and approach the barrier at frequency $f = \omega_{\mathrm{eff}}/2\pi$.
Using the Wentzel–Kramers–Brillouin (WKB) approximation, the transmission probability through the tunnel barrier is given by \cite{shankar2011,karnakov2013,griffiths_2005} 

\begin{equation} 
\Tilde{T} = \mathrm{exp}\bigg(-2\int_{y_{1}}^{y_{2}}\kappa(y)\mathrm{d}y\bigg),
\label{eq:transmission probability}
\end{equation}
where $y_{1}$, $y_{2}$, are the two classical turning points for the potential barrier, $\kappa(y) = \sqrt{2m(U_{\mathrm{tot}}(y) - E)}/\hbar$, and $U_{\mathrm{tot}}(y)$ is the form of the barrier in the total potential energy curve.\\

The average tunneling-limited lifetime of a trapped atom is then defined by

\begin{equation} 
\tau_{\mathrm{tun}}(y_{0}) = \frac{1}{\Gamma_{\mathrm{tun}}(y_{0})}.
\label{eq:tunnelling loss lifetime}
\end{equation}

\begin{figure}[ht]
\includegraphics[width = \linewidth]{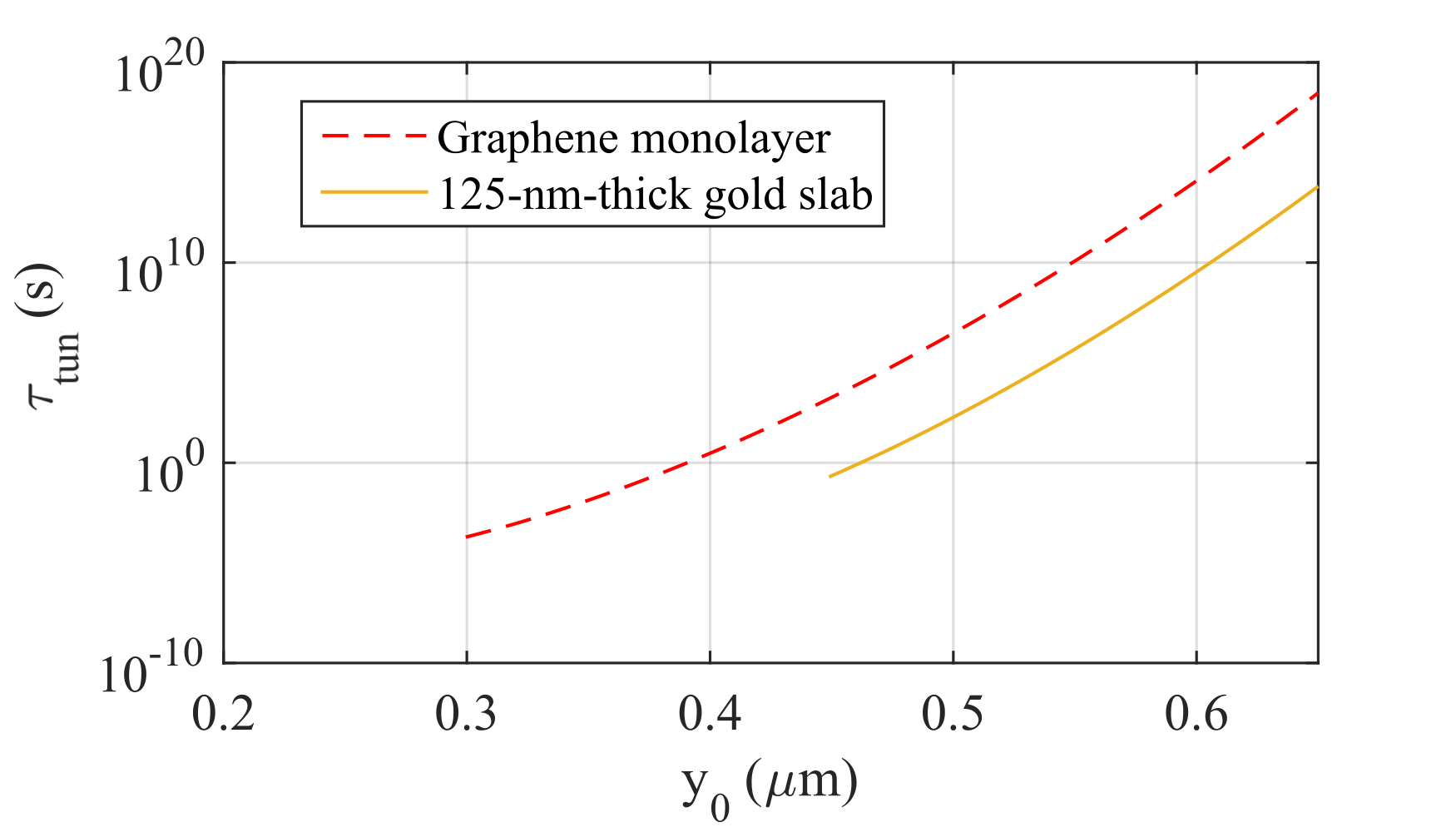}
\caption{Tunneling lifetime, $\tau_{\mathrm{tun}}$, calculated versus the position of the harmonic trap center, $y_{0}$, for an \textsuperscript{87}Rb atom trapped near a graphene monolayer (dashed red curve) and a $125$-$\SI{}{\nano\metre}$-thick gold slab (solid yellow curve). The weaker CP attraction for graphene gives rise to a higher, wider, tunnel barrier and, consequently, a higher tunneling lifetime. Parameters: $T = \SI{300}{\kelvin}$, $\omega_{r} = 2\pi\times\SI{20}{kHz}$.}
\label{fig:Tunnel_lifetime_main}
\end{figure}
We calculated the tunneling loss rates for our model systems using the Gamow formalism described above. Figure \ref{fig:Tunnel_lifetime_main} shows the resulting lifetimes, $\tau_{\mathrm{tun}}$, calculated versus the position of the trap center, $y_{0}$ from a graphene monolayer (dashed red curve) and a thin gold slab (solid yellow curve). The minimum distance that atoms can be trapped from the surface is marked by the left-hand ends of the two curves, where the tunnel barrier vanishes. Comparison of the curves shows that using a graphene monolayer reduces this distance to $\sim \SI{0.3}{\micro\metre}$ compared to the value of $\sim\SI{0.45}{\micro\metre}$ for the gold layer. For $y_{0} > \SI{0.5}{\micro\metre}$, the tunneling lifetime for the single layer of graphene is orders of magnitude higher than for the gold slab due to the weaker CP potential.

\subsection{Atom losses due to Johnson Noise}
\label{sec:Temporal_Johnson Noise}

Magnetically trapped atoms only remain trapped when they are in a low field seeking state with the magnetic moment aligned anti-parallel to the direction of the magnetic field. In order to keep $m_{F}$ a good quantum number, an offset magnetic field of a few 
Gauss is typically maintained at the trapping position in atom-chip experiments \cite{Lin2004}. Given a Zeeman splitting of, for 
example, 0.7 MHz/G for the $^{87}$Rb ground state, this produces 
transition frequencies of a few MHz between hyperfine states with different $m_{F}$ values, thus making the trapped atoms susceptible to magnetic fields in that frequency range and therefore to noise in the radio frequency domain.\\  

Johnson noise arises from electrical noise currents within a conductor, which produce fluctuations of the magnetic field \cite{henkel1999}. For near-surface traps formed between $\approx \SI{1}{\micro\metre}$ and 1 mm from a metallic conductor on a conventional atom-chip, Johnson noise is usually the main limitation on the lifetime of the atom cloud \cite{Lin2004,Jones2003,Harber2003}. For example, the measured lifetime of atoms trapped $\approx \SI{1}{\micro\metre}$ from thick metallic conductors is limited to only $\approx 0.1$ s by the effects of Johnson noise \cite{Lin2004}.\\

The key advantage of using trapping wires made from graphene, or other two-dimensional conductors is their reduced level of Johnson noise \cite{SinucoA, SinucoB}. This advantage originates from their orders-of-magnitude lower sheet electron density compared with metals, which dominates over the tendency of higher carrier mobility to increase current fluctuations. We now explore this advantage using two models of increasing sophistication. Firstly, a crude estimate based on previous models for metallic conductors and derived using the fluctuation-dissipation theorem \cite{Lin2004,henkel1999,henkel2001}. Secondly, we present a rigorous quantum field theoretical calculation for 2D materials involving the full Green's function for the system, determined from the reflection coefficients for graphene and 2D multilayers.\\

Comparing graphene and gold wires with a given top surface area, $A$, the ratio of the number of free electrons in graphene, $N_{G}$, to that in gold, $N_{Au}$, is $N_{G}/N_{Au} = n_{G}/n_{Au}t_{Au}$, where $t_{Au}$ is the thickness of the gold wire and $n_{G}$, $n_{Au}$ are, respectively, the sheet and volume electron densities of undoped graphene and gold. Taking $n_{Au} = \SI{5.9e28}{\metre^{-3}}$ for bulk gold and typical values of $t_{Au} = \SI{1}{\micro\metre}$ and $n_{G} = \SI{9e14}{\metre^{-2}}$ \cite{fang_konar_2007} gives $N_{G}/N_{Au} = 1.5 \times 10^{-8}$. 
The carrier mobility of gold is $\mu_{Au}\approx 4.3 \times 10^{-3}$ m$^2$/Vs and, in graphene, mobilities up to $\mu_G\approx 20 $ m$^2$/Vs have been reported in free-standing membranes \cite{bolotin2008, chen_jang_2008}. For graphene on a substrate, the electron mobility is typically at least an order of magnitude lower, leading to the estimate $\mu_{Au}/\mu_G \gtrsim 2 \times 10^{-4}$.\\
 
These results allow us to anticipate that the Johnson noise will be far smaller in graphene than in gold. Indeed, to make a rough initial estimate of this intuitive advantage, we now use the model presented in \cite{Lin2004,henkel1999,henkel2001} to evaluate the expected spin-flip lifetime enhancement. For an atom trapped at distance, $d$, from a metal film of width $w \gg t$ and resistivity $\rho$ at temperature $T$, the $\vert F,m \rangle \rightarrow \vert F,m-1 \rangle$ spin-flip rate, given in $\SI{}{\second^{-1}}$, is $\Gamma = C(T/\rho)\times[d(1+d/t)(1+2d/w)]^{-1}$, where $C$ is a constant that depends on the Clebsch-Gordon coefficient for the transition and on the transition frequency \cite{Lin2004,henkel1999,henkel2001}. Assuming, as a crude initial approximation, that this formula can also be used to estimate the rate of spin flips induced by electrons in graphene, the ratio of the lifetimes, $\tau_G$ and $\tau_{Au}$, of atom clouds trapped at a distance $d$ above graphene and gold wires, respectively, is

\be \label{eq:ratio}
\frac {\tau_G}{\tau_{Au}}=\frac{\Gamma_{Au}}{\Gamma_G}=\frac{\rho_{G}}{\rho_{Au}}\frac{\left(1+\frac{d}{t_G}\right)}{\left(1+\frac{d}{t_{Au}}\right)},
\ee
where $t_G=0.345$ nm is the thickness of a graphene monolayer and $\rho_{Au}=1/(n_{Au}e\mu_{Au})$, $\rho_G=t_G/(n_G e\mu_G)$ are the resistivities of the gold and graphene
and $e$ is the electron charge.\\

Since $d \gg t_G$, it follows that\\

\be
\begin{split}
\frac {\tau_G}{\tau_{Au}} \approx \left(\frac{n_{Au}}{n_G}\right)\left(\frac{\mu_{Au}}{\mu_G}\right)\frac{t_{Au}d}{\left(d+t_{Au}\right)}\\
= \left(\frac{N_{Au}}{N_G}\right)\left(\frac{\mu_{Au}}{\mu_G}\right)\frac{d}{\left(d+t_{Au}\right)}.
\end{split}
\ee

Unless $d \ll t_{Au}$, which is not the case for presently-attainable trapping distances, the final term in the above equation is of order unity and so $\tau_G/ \tau_{Au}$ depends primarily on the relative number of free electrons in gold and in graphene and on their mobility ratio.\\

Using the values for the carrier mobility and density given above we arrive at $\tau_G/\tau_{Au} \gtrsim 1.3 \times 10^4 d/(d+t_{Au})$. 
We thus predict that for atoms trapped $\approx \SI{1}{\micro\metre}$ away from a conductor, the lifetime will increase from $\approx 0.1$ s for a 1-$\SI{}{\micro\metre}$-thick metallic wire, similar to that reported in \cite{Lin2004}, to $\gtrsim 600$ s for graphene, i.e. an increase by a factor of $\tau_G/\tau_{Au} \gtrsim 6.3 \times 10^3$. The physical reason for this is that although electrons in graphene have a higher mobility than in a metal, and so produce more Johnson noise per carrier, this is more than compensated by the far lower number of charge carriers in the graphene.\\

In the next section, we derive an expression for the Johnson noise produced by van der Waals heterostructures. To quantify the advantages of using 2D conductors, rather than metal wires, to reduce noise in atom-chips we consider the particular case of graphene conduction channels. However, similar advantages are expected from other 2D materials due to their low number of electric current carriers.  

\subsubsection{Transition rates in terms of dyadic Green's functions}
The magnetic moment vector associated with the transition $\ket{i} \to \ket{f}$ of an atom is given by \cite{Buschow2013, Rekdal2004}

\begin{equation}
\boldsymbol{\mu} = -\bra{i}\frac{\mu_{B}}{\hbar}\Big(g_{S}\hat{\mathbf{S}} + g_{L}\hat{\mathbf{L}} - g_{I}\frac{m_{\mathrm{e}}}{m_{\mathrm{nuc}}}\hat{\mathbf{I}}\Big)\ket{f},
\label{eq:magnetic moment}
\end{equation}
where $\hat{\mathbf{S}}$, $\hat{\mathbf{L}}$, and $\hat{\mathbf{I}}$ are the electron spin operator, the electron orbital angular momentum operator, and the total nuclear angular momentum operator, respectively, with their corresponding Land\'{e} g-factors $g_{S}$, $g_{L}$, and $g_{I}$, $m_{\mathrm{e}}$ is the electron mass and $m_{\mathrm{nuc}}$ is the nuclear mass. Here, the magnitude of the angular momentum, for example, $\hat{\mathbf{S}}$, is $\sqrt{S(S+1)}\hbar$ and the eigenvalue of the $z$-component of $\hat{\mathbf{S}}$, i.e. $\hat{\mathbf{S}_{z}}$, is $m_{S}\hbar$, where $S$ and $m_{S}$ are the corresponding quantum numbers for $\hat{\mathbf{S}}$ and $\hat{\mathbf{S}_{z}}$, respectively.\\

Taking $L = 0$ for the electronic ground-state and neglecting the term containing the total nuclear angular momentum operator $\hat{\mathbf{I}}$ in equation \eqref{eq:magnetic moment} because $m_{\mathrm{e}} \ll m_{\mathrm{nuc}}$, the magnetic moment vector becomes $\boldsymbol{\mu} = -\mu_{B}g_{S}\bra{i}\hat{\mathbf{S}}\ket{f}/\hbar$, where $g_{S} = 2$, and the rate of magnetic spin-flip transitions from an initial hyperfine magnetic state $\ket{i}$ to another state $\ket{f}$ is given by \cite{Rekdal2004}

\begin{multline}
\Gamma_{\mathrm{JN}} = \mu_{0}\frac{2(\mu_{B}g_{S})^{2}}{\hbar^{2}}\sum_{j,k}\Big\{\bra{f}\hat{\mathbf{S}}_{j}\ket{i}\bra{i}\hat{\mathbf{S}}_{k}\ket{f}\\
\times \mathrm{Im}[\curl{\curl{\mathbf{G}(\mathbf{r}_{0}, \mathbf{r}_{0}, \omega_{if})}}]_{jk}(\bar{n}_{\mathrm{th}}+1)\Big\},
\label{eq:Splin-flip rate}
\end{multline}
where $\hat{\mathbf{S}}_{j,k}$ denotes the $j$ and $k$ components of the electron spin operator $\hat{\mathbf{S}}$, and $\mathbf{G}(\mathbf{r}_{0}, \mathbf{r}_{0}, \omega_{if})$ is the total dyadic Green's function describing the electromagnetic field of the transition frequency $\omega_{if}$ at $\mathbf{r}_{0}$ due to a \emph{magnetic} dipole located at $\mathbf{r}_{0}$ (see Appendix \ref{supplement: Green's function}). The mean thermal photon occupation number is given by

\begin{equation}
\bar{n}_{\mathrm{th}} = \frac{1}{\mathrm{e}^{\hbar\omega_{if}/k_{B}T} - 1},
\label{eq:mean_thermal occupation}
\end{equation}
where $T$ is the temperature of the electromagnetic field system that causes the spin-flip transitions, rather than of the trapped atoms, and $\omega_{if}$ is the angular frequency of the radiation due to magnetic spin-flip transitions. The Johnson noise lifetime of a single atom is defined as
\begin{equation}
\tau = \frac{1}{\Gamma_{\mathrm{JN}}}.
\label{eq:Lifetime}
\end{equation}

Note that, mathematically, $\mathbf{G}(\mathbf{r}_{0}, \mathbf{r}_{0}, \omega_{if})$ can be written as the sum of a Green's tensor, describing the field due to a dipole in an infinitely extended homogeneous bulk medium, vacuum for example, and a scattering Green's tensor describing the reflected field in the presence of reflective bodies, so that

\begin{equation}
\mathbf{G}(\mathbf{r}_{0}, \mathbf{r}_{0}, \omega_{if}) = \mathbf{G}^{(0)}(\mathbf{r}_{0}, \mathbf{r}_{0}, \omega_{if}) + \mathbf{{G}}^{(1)}(\mathbf{r}_{0}, \mathbf{r}_{0}, \omega_{if}). 
\label{eq:Total_green_function}
\end{equation}

The explicit forms of $\mathbf{G}^{(0)}(\mathbf{r}_{0}, \mathbf{r}_{0}, \omega_{if})$ can be found in \cite{Novotny2006, Buhmann_ii} and are summarized in Appendix \ref{supplement: Green's function}. 
Owing to the general properties of a Green's tensor, we have 

\begin{equation}
\mathrm{Im}[\curl{\curl{\mathbf{G}(\mathbf{r}, \mathbf{r}_{0}, \omega)}}] = \frac{\omega^{2}}{c^{2}}\mathrm{Im}[\epsilon(\omega)\mu(\omega)\mathbf{G}(\mathbf{r}, \mathbf{r}_{0}, \omega)],
\label{eq:Green function properties}
\end{equation}
where $\epsilon(\omega)$ and $\mu(\omega)$ are, respectively, the permittivity and permeability of the medium in which the field and source points are located. 
The imaginary part of the Green's tensor in vacuum has a simple form:

\begin{equation}
\mathrm{Im}[\mathbf{G}^{(0)}(\mathbf{r}_{0}, \mathbf{r}_{0}, \omega)]_{jk} = \frac{1}{6\pi}\frac{\omega}{c}\delta_{jk},
\label{eq:imaginary part of vaccum Green}
\end{equation}
where $\delta_{jk}$ is the Kronecker delta, which allows us to determine the spin-flip rates in vacuum. 

\subsubsection{Johnson noise lifetime calculation}
In this section, we use Eq. \eqref{equ:MagPot} to determine an appropriate value of the atomic transition frequency for input into our Johnson noise lifetime calculations in the presence of the magnetic trapping field.
For the $\ket{F, m_{F}} = \ket{2, 2}$ $5^{2}S_{1/2}$ ground-state of the \textsuperscript{87}Rb atoms considered here, $g_{F} = 1/2$ \cite{Steck2020, Buschow2013}. Here, we only consider the Zeeman transition from $\ket{2, 2}$ to $\ket{2, 1}$ to facilitate direct comparison with the results of Ref. \cite{Lin2004}. The angular frequency of the radiation is then given by 
\begin{equation}
\omega_{if} = \frac{\mu_{B}\abs{\mathbf{B(\mathbf{r}_{0})}}}{2\hbar}.
\label{eq:omega_spin_flip}
\end{equation}
Taking $\abs{\mathbf{B(\mathbf{r}_{0})}} = \SI{0.8e-4}{\tesla}$ gives $\omega_{if} = 2\pi\times\SI{560}{kHz}$.
Comparing with the hyperfine splitting frequency for the ground-state of the \textsuperscript{87}Rb atom, which is $2\pi\times\SI{6.83}{GHz}$, we now see that our assumption that $m_F$ is a good quantum number is justified. The method for calculating the Clebsch-Gordon coefficients associated with the spin-flip transition matrix elements, $\bra{f}\hat{\mathbf{S}}_{j,k}\ket{i}$, can be found in \cite{Zettili2009}. For completeness, we note that these matrix elements are
\begin{equation} \label{eq: Clebsch-Gordon coeff}
\begin{split}
\bra{2,2}\hat{\mathbf{S}}_{x}\ket{2,1} &=\phantom{-}\bra{2,1}\hat{\mathbf{S}}_{x}\ket{2,2} = \frac{1}{4},\\
\bra{2,2}\hat{\mathbf{S}}_{y}\ket{2,1} &=-\bra{2,1}\hat{\mathbf{S}}_{y}\ket{2,2} = \frac{\mathrm{i}}{4},\\
\bra{2,2}\hat{\mathbf{S}}_{z}\ket{2,1} &=\phantom{-}\bra{2,1}\hat{\mathbf{S}}_{z}\ket{2,2} = 0. \end{split}
\end{equation}

\begin{figure}[ht]
\includegraphics[width = \linewidth]{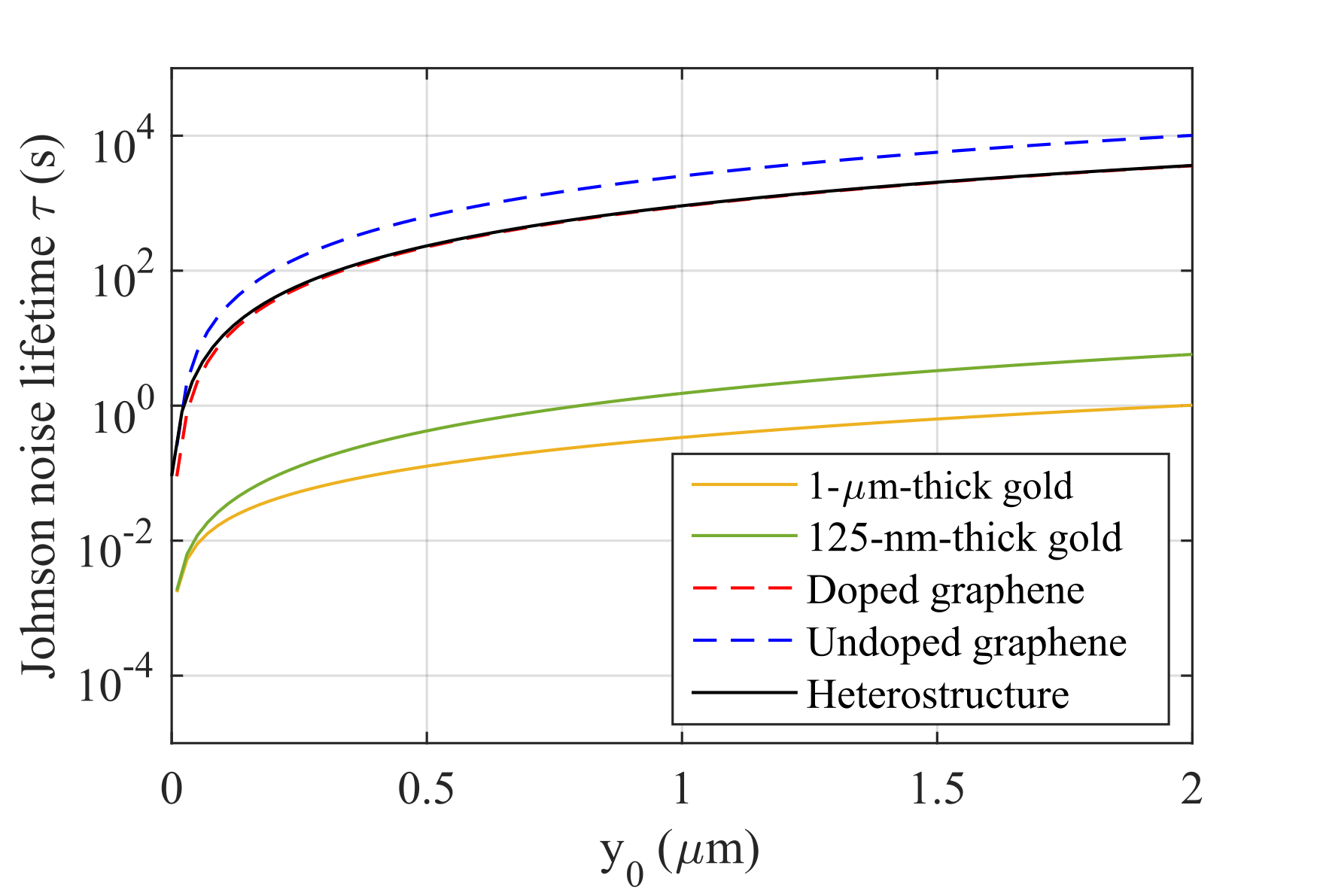}
\caption{Johnson noise lifetimes, $\tau$, calculated (using Eqs. \eqref{eq:Splin-flip rate} and \eqref{eq:Lifetime}) versus the position of the harmonic trap center, $y_{0}$, for an ultra-cold gas of \textsuperscript{87}Rb atoms trapped above: an undoped graphene monolayer (dashed blue curve); a doped graphene monolayer with Fermi energy $E_{F} = \SI{0.1}{eV}$ (dashed red curve); an hBN-encased graphene-based heterostructure (solid black curve); a $1$ $\SI{}{\micro\metre}$-thick gold slab (solid yellow curve); a $125$ $\SI{}{\nano\metre}$-thick gold slab (solid green curve). The graphene monolayers yield orders of magnitude longer lifetimes than the $1$ $\SI{}{\micro\metre}$-thick gold slab because they have far lower electromagnetic reflectance than gold (see text). Parameters: $T = \SI{300}{\kelvin}$, $\omega = 2\pi\times\SI{560}{kHz}$.}
\label{fig:Johnson_Gr_gold}
\end{figure}

To proceed with our calculations of the transition rates and comparisons for different surface materials, we consider the typical thickness of the metallic wires used to generate the magnetic field in atom-chips, which is $\sim \SI{1}{\micro\metre}$ \cite{Lin2004}. Figure \ref{fig:Johnson_Gr_gold} shows the Johnson noise-limited lifetimes of the atom cloud, $\tau$, calculated versus atom-surface distance, $y_{0}$, for a $1$-$\SI{}{\micro\metre}$-thick gold slab (solid yellow curve), a $125$-$\SI{}{\nano\metre}$-thick gold slab (solid green curve), a doped graphene monolayer with $E_{F} = \SI{0.1}{eV}$ (dashed red curve), an undoped graphene monolayer (dashed blue curve) and a heterostructure consisting of a graphene monolayer encased by two 10-nm-thick hBN layers (solid black curve). As in \cite{Lin2004}, these Johnson noise lifetimes are calculated using Eqs. (\ref{eq:Splin-flip rate}) and (\ref{eq:Lifetime}). The graphene monolayers yield far longer lifetimes than the gold wires do, even for a small wire thickness of $\SI{125}{\nano\metre}$. Making gold wires thinner than $125$ $\SI{}{\nano\metre}$ is possible, but their resistivities then become higher than for bulk gold \cite{salem_japha2010}. At $y_{0} = \SI{1}{\micro\metre}$, the lifetimes for the undoped graphene monolayer and the $1$ $\SI{}{\micro\metre}$-thick gold slab are $\sim \SI{2500}{\second}$ and $\SI{0.34}{\second}$, respectively, giving a lifetime ratio of $\sim \SI{7.4e3}{}$, which is broadly consistent with the estimate of $\sim \SI{6.3e3}{}$ obtained from Eq. \eqref{eq:ratio}. The lifetime for the heterostructure is slightly longer than that for the doped graphene layer.  

We conclude that Johnson noise in graphene conductors will produce negligible spin-flip losses compared to the thick ($t_{Au}\sim \SI{1}{\micro\metre}$) metal wires typically used in atom-chips, where it dominates the loss rate. Consequently, our analysis of graphene atom-chips will, henceforth, focus on the effects of tunneling and three-body losses and of spatial imperfections. We note, however, from Eq. (\ref{eq:ratio}) that the lifetime above metallic conductors can be increased by decreasing their thickness $t_{Au}$ and, hence, $N_{Au}$. For wires with $t_{Au}=125$ nm, $\tau_G/\tau_{Au} \sim 1400$. Taking the limit of the gold layer thickness to its lattice constant, 0.4 nm, gives $\tau_G/\tau_{Au} \sim 5$. So the advantage of graphene over metallic conductors persists even if the metal wire could be thinned close to the theoretical limit of a monolayer. To our knowledge, though, graphene and other exfoliated van der Waals materials are the only monolayers so far produced. Moreover, their hexagonal crystal structure and resulting light-like linear energy band dispersion relations ensure that they can carry high currents despite their low thickness and carrier density. However, if high-quality metallic monolayers could be produced, their low electron density may reduce the Johnson noise and Casimir-Polder potential to levels comparable with exfoliated 2D materials. 

\subsection{Negligible corrugation effects}
Spatial meandering of the current stream lines can, in principle, be created in four ways: deviation from strictly two-dimensional current flow (analogous to surface roughness of 3D conductors); edge roughness resulting from imperfect lithography; electrons scattering from one another or from phonons; spatial variations in the electron potential energy created by impurities or imperfections in, or near, the conducting channel \cite{SinucoA,SinucoB,kruger_andersson_2007, schumm_esteve_2005, wang_lukin_2004}. We now consider the importance of each potential source of roughness in turn.\\

When graphene is encapsulated in hBN, surface roughness and non-two-dimensionality in graphene is only of order 12 pm \cite{thomsen_gunst_2017} because the hBN provides an ultraflat surface for the graphene and is closely lattice matched to it \cite{dean_young_meric_2010, decker_wang_2011, xue_sanchez_2011}. Such low roughness is consistent with that of an individual graphene layer in bulk graphite and will have a negligible effect on the atom trapping potential landscape. 
Edge roughness will be determined by the quality of the lithography used to define and create the conducting channels. Since graphene is two dimensional, there will be negligible vertical fluctuations in the channel wall. Edge fluctuations along the channel will be determined by the lithographic process used and comparable to those in existing atom-chips with metallic conductors. For electron beam lithography, the edge fluctuations will be of order 35 nm \cite{xu_lee_2016}, whereas for helium ion beams, values below 5 nm are attainable \cite{aigner_pietra_2008}.\\

In metallic conductors, grain boundaries give rise to local electron scattering processes, which can be detected via their effect on the current flow pattern and resulting modulation of the trapping potential and BEC atom density \cite{SinucoA,SinucoB,aigner_pietra_2008,japha_entin_2008}. By contrast, graphene monolayers contain no grains to induce position-specific scattering processes and resulting atom density fluctuations. Electron-electron and electron-phonon scattering events do occur, but these are spatio-temporally stochastic, rather than occurring at particular fixed positions within the conducting channel and will therefore not produce roughness in the trapping potential and BEC density profile because of time averaging. Moreover, since their characteristic length scales are shorter than the typical dimensions of atom-chip wires, ballistic transport effects do not need to be considered. However, electron scattering mechanisms do affect the diffusive electron mobility and, hence, the Johnson noise-limited spin-flip lifetime of the trapped atom cloud.\\

Spatial fluctuations in the electronic potential energy created by imperfections and impurities that are either within the graphene or accumulate at interfaces in hBN-encased graphene structures have been studied theoretically and measured in resonant-tunneling experiments \cite{decker_wang_2011, britnell_2013, martin_akerman_2007, li_hwang_rossi_2011, yan_fuhrer_2011, yankowitz_xue_cormode_2012}. Self-consistent calculations \cite{li_hwang_rossi_2011, yan_fuhrer_2011}, which give excellent quantitative agreement with measurements of graphene's electron mobility, $\mu_G$, versus impurity density and with scanning probe surface studies \cite{martin_akerman_2007}, predict that the correlation length of these potential fluctuations is $\approx 10$ nm. Recent experiments on graphene-boron nitride tunnel transistors have shown that for graphene monolayers encased by several layers of hexagonal boron nitride (hBN), the correlation length is $\approx 12$ nm \cite{britnell_2013,greenaway_vdovin_mishchenko_2015}. Consequently, the associated small-angle current meander will have negligible effect on the potential landscape of atoms trapped even as close as 150 nm from the graphene and will therefore not influence the minimum atom-surface trapping distance.\\

When graphene is placed or grown epitaxially on hBN, the small lattice mismatch between the two materials gives rise to a strain-induced moir\'{e} pattern and superlattice potential, which can modify the electronic properties of electrons within the graphene. Moir\'{e} periods up to 80 nm have been realized \cite{davies_albar_2017} and further increases in period may modulate the current flow on a length scale long enough to produce detectable variation in the density profile of a BEC trapped nearby. Such variations could yield information about the superlattice potential and the underlying strain mechanisms.  

\section{Lifetime of A tRapped atomic BEC}

In this section, we find an analytical expression for the total loss rate of an elongated atomic BEC trapped in the vicinity of an atom-chip. We consider contributions from atom tunneling towards the chip surface (Sec. \ref{sec:tunnelling towards the chip surface}), Johnson-noise induced losses (Sec. \ref{sec:Temporal_Johnson Noise}) and the 3-body loss mechanism. 

\subsection{Methodology}
First, we consider a harmonic magnetic trapping field, $\mathbf{B}(\mathbf{r})$, formed near the surface of an atom-chip in the coordinate system shown in Fig.~\ref{fig:chip_diagram}, where $\mathbf{r} = (x,y,z)$, $\omega_{x}$, $\omega_{y}$, and $\omega_{z}$ are the characteristic trapping frequencies in the $x$-, $y$-, and $z$-axes, respectively, and the trap center is located at $\mathbf{r}_{c} = (0,y_{c},0)$. The potential energy profile of an atom interacting with this magnetic field is modelled by an anisotropic three-dimensional harmonic-oscillator potential  
\begin{equation}
U(x, y, z) = \frac{1}{2}m\big(\omega_{x}^{2}x^{2} + \omega_{y}^{2}(y-y_{c})^{2} + \omega_{z}^{2}z^{2}\big).\\
\label{eq:U_Harmonic_main_text}
\end{equation} 
Note that this magnetic potential originates from the interaction of the magnetic moment of the trapped atom and the magnetic field given in Eq. \eqref{equ:MagPot} and that the actual potential profile of an atom-chip trap is determined by the wire and current configurations. Equation \eqref{eq:U_Harmonic_main_text} gives a good approximation for the potential landscape generated by the Z-shaped trapping wires often used in atom-chip experiments.\\

We assume that such a magnetic trap has cylindrical symmetry and is elongated along the $z$-axis, so that $\omega_{r} = \omega_{x,y}$, and $\omega_{r} \gg \omega_{z}$, where $\omega_{r}$ denotes the trapping frequency in the radial direction (i.e. in the $x$-$y$ plane). We also assume that $\omega_{r}$ is so high that the trapped atoms only occupy the ground-state energy in the radial direction. To include the perturbing effect of the CP potential on the effective trapping frequency in the $y$-direction, henceforth we approximate the radial trapping frequency as $\omega_{r} = \sqrt{\omega_{x}\omega_{\mathrm{eff}}}$. An additional offset magnetic field, $\mathbf{B}_{0} = (0, 0, B_{z})$, of order $\SI{}{mT}$, is added in the $z$-direction to ensure that the magnetic field is non-zero at the trap center. Together, these assumptions enable us to treat the magnetic potential energy landscape as a highly elongated, quasi one-dimensional, trap.\\

It follows from the above assumptions about the trapping frequencies that the chemical potential, $\mu$, of the condensate must satisfy the following constraints 

\begin{equation} 
5\hbar\omega_{z} < \mu < \frac{3}{2}\hbar\omega_{r},
\label{eq:chemical_potential_condition}
\end{equation} 
which allows us to further assume that the mean atom density profile of the condensate can be described by a one-dimensional Thomas-Fermi distribution in the elongated ($z$) direction, and by the Gaussian ground-state wave function of a quantum harmonic oscillator in the tightly-confining radial ($r$) direction \cite{pethick_smith_2008}. The atom density profile is then given by 

\begin{equation} 
\rho_{0}(r, z) = \frac{1}{U_{0}}\Big(\mu_{\mathrm{eff}} - \frac{m\omega_{z}^{2}}{2}z^{2}\Big)e^{-r^{2}/2a_{r}^2},
\label{eq:Thomas-Fermi distribution_main_text}
\end{equation} 
where $U_{0} = 4\pi\hbar^{2}a_{T}/m$, is the effective interaction strength for a pair of slowly moving atoms of s-wave scattering length $a_{T}$ \cite{olshanii_1998}, $\mu_{\mathrm{eff}} = \mu - \hbar\omega_{r}$, $m = \SI{1.44e-25}{kg}$ is the mass of an \textsuperscript{87}Rb atom, $r = \sqrt{x^{2} + (y-y_{c})^{2}}$ is the radial distance relative to the center of the trap, and $a_{r} = \sqrt{\hbar/m\omega_{r}}$ is the characteristic harmonic oscillator length.\\

Integrating Eq. \eqref{eq:Thomas-Fermi distribution_main_text} over the radial co-ordinate gives the mean line density profile along the $z$-axis (see Appendix \ref{sec:3b_loss_appendix})

\begin{equation}
n_{0}(z) = \frac{2\pi a_{r}^{2}}{U_{0}}\Big(\mu_{\mathrm{eff}} - \frac{m\omega_{z}^{2}}{2}z^{2}\Big),
\label{eq:line_density_main_text}
\end{equation}
where, the chemical potential $\mu$ of the trapped atom cloud is determined by the peak mean line density, i.e. at $z = 0$, as follows \cite{menotti_stringari_2002}:

\begin{equation}
\mu = \big(2a_{T}n_{0}(0) + 1\big)\hbar\omega_{r},
\label{eq:chemical_potential_expression_line_den}
\end{equation}
where $a_{T} = \SI{5.6}{\nano\metre}$ is the scattering length for \textsuperscript{87}Rb atoms in the $\ket{F,m_{F}} = \ket{2,2}$ state \cite{roberts_1998}.\\

\begin{figure}[ht]
\centering
\includegraphics[width = \linewidth]{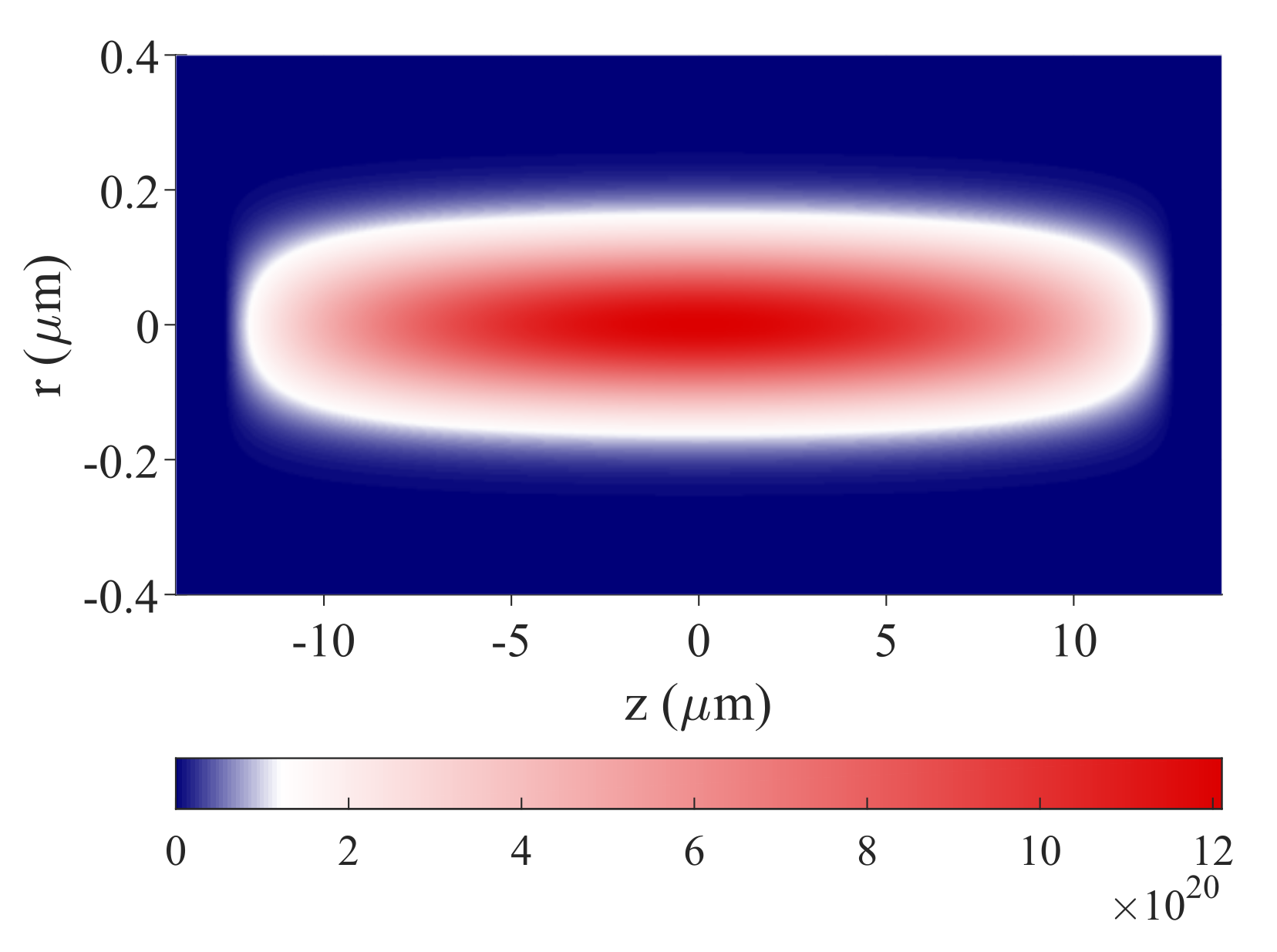}
\caption{Color map of the atom volume density calculated for the trapped atom cloud using the Thomas-Fermi distribution in Eq. \eqref{eq:Thomas-Fermi distribution_main_text}. The color bar scale is in units of $\SI{}{m^{-3}}$. Parameters: $\omega_{r} = 2\pi\times\SI{20}{kHz}$, $\omega_{z} = 0.006\times\omega_{r}$, $N = 750$.}
\label{CP_many_structures}
\end{figure}

We now define the total lifetime of the trapped atom cloud, $\tau_{\mathrm{tot}}$, to be the time taken for the initial peak atom line density, $n_0(z = 0)$, to drop below the smallest experimentally-detectable line density, which we take to be $n_{\mathrm{min}} = \SI{3e6}{m^{-1}}$ \cite{smith_aigner_2011}. We determine the upper limit of $\tau_{\mathrm{tot}}$ by taking $n_0(z = 0)$ to be the maximum possible value, $n_{\mathrm{max}}=14.8\times n_{\mathrm{min}}$, satisfying inequality \eqref{eq:chemical_potential_condition}.\\

Let us now consider the density-dependent loss rates originating from three distinct atom loss mechanisms; Johnson noise-induced spin flips, quantum tunneling to the chip surface, and three-body processes. The Johnson noise-induced loss rate is

\begin{equation}
\frac{\mathrm{d}n_{0}(z)}{\mathrm{d}t}\biggr\rvert_{\mathrm{JN}} = -\Gamma_{\mathrm{JN}} n_{0}(z),
\label{eq:Johnson noise induced loss rate}
\end{equation}
where $\Gamma_{\mathrm{JN}}$ is the Johnson noise-induced spin-flip transition rate given in Eq. \eqref{eq:Splin-flip rate}.

As described above, the tunneling loss rate has a similar form

\begin{equation}
\frac{\mathrm{d}n_{0}(z)}{\mathrm{d}t}\biggr\rvert_{\mathrm{tun}} = -\Gamma_{\mathrm{tun}} n_{0}(z),
\label{eq:tunnelling loss rate}
\end{equation}
where $\Gamma_{\mathrm{tun}} = \omega_{\mathrm{eff}}\Tilde{T}/2\pi$.\\

By contrast, the three-body loss rate is proportional to the cube of the mean line density \cite{bouchoule_schemmer_henkel_2018,schemmer2018}, 

\begin{equation} 
\frac{\mathrm{d}n_{0}(z)}{\mathrm{d}t}\biggr\rvert_{\mathrm{3b}} = -\Gamma_{\mathrm{3b}}n_{0}(z)^{3},
\label{eq:1D loss rate_main_text}
\end{equation}
where $\Gamma_{\mathrm{3b}}  = \kappa_{\mathrm{Rb}}/12 \pi^{2} a_{r}^{4}$ and $\kappa_{\mathrm{Rb}} = \SI{1.8e-41}{m^{6}s^{-1}}$ is the three-body recombination rate for \textsuperscript{87}Rb in the $F = m_{F} = 2$ state \cite{soding1999}.\\

Combining all three distinct loss rates gives the total loss rate 
\begin{equation} 
\frac{\mathrm{d}n_{0}(z)}{\mathrm{d}t}\biggr\rvert_{\mathrm{tot}} = -\Gamma_{\mathrm{3b}}n_{0}(z)^{3} - (\Gamma_{\mathrm{tun}}+\Gamma_{\mathrm{JN}})n_{0}(z).
\label{eq:total loss rate_main_text}
\end{equation}

In this paper, we will only consider losses occurring at $z = 0$, where the line density peaks and so the total loss rate is maximal. Hence, we determine the lower limit on the total lifetime given by the integral:

\begin{equation} 
\tau_{\mathrm{tot}} = \int_{n_\mathrm{max}}^{n_{\mathrm{min}}}\frac{\mathrm{d}n_{0}(z)}{-\Gamma_{\mathrm{3b}}n_{0}(z)^{3} - (\Gamma_{\mathrm{tun}}+\Gamma_{\mathrm{JN}})n_{0}(z)}\Big\rvert_{z = 0},
\label{eq:total life time}
\end{equation}
which can be integrated analytically to yield:

\begin{equation}
\tau_{\mathrm{tot}} = \frac{\log{\Bigg[\frac{\Gamma_{\mathrm{3b}}n_\mathrm{min}^{2} + (\Gamma_{\mathrm{tun}}+\Gamma_{\mathrm{JN}})}{\alpha^{2}\Gamma_{\mathrm{3b}}n_\mathrm{min}^{2} + (\Gamma_{\mathrm{tun}}+\Gamma_{\mathrm{JN}})}\Bigg]}\\
+ 2\log{(\alpha)}}{2(\Gamma_{\mathrm{tun}}+\Gamma_{\mathrm{JN}})},
\label{eq:total_life_time_analytic}
\end{equation} 
where $\alpha = n_{\mathrm{max}}/n_{\mathrm{min}} = 14.8$.

\subsection{Results}
In this section, we calculate and compare atom-cloud lifetimes for three different surface structures: a $1$ $\SI{}{\micro\metre}$-thick gold slab, a graphene monolayer, and a graphene monolayer encased by two $10$ $\SI{}{\nano\metre}$-thick hBN layers. The first structure is representative of the present generation of atom-chips, and provides typical lifetimes for comparison with graphene-based devices. The second structure is the theoretical limit for 2D materials and exemplifies the predicted improvements in performance and functionality. The third atom-chip structure is within the scope of existing fabrication techniques for vdW heterostructures. The graphene conductor is encased within hBN multilayers, which support it and shield it from adsorbates.\\

\begin{figure}[ht]
\includegraphics[width = \linewidth]{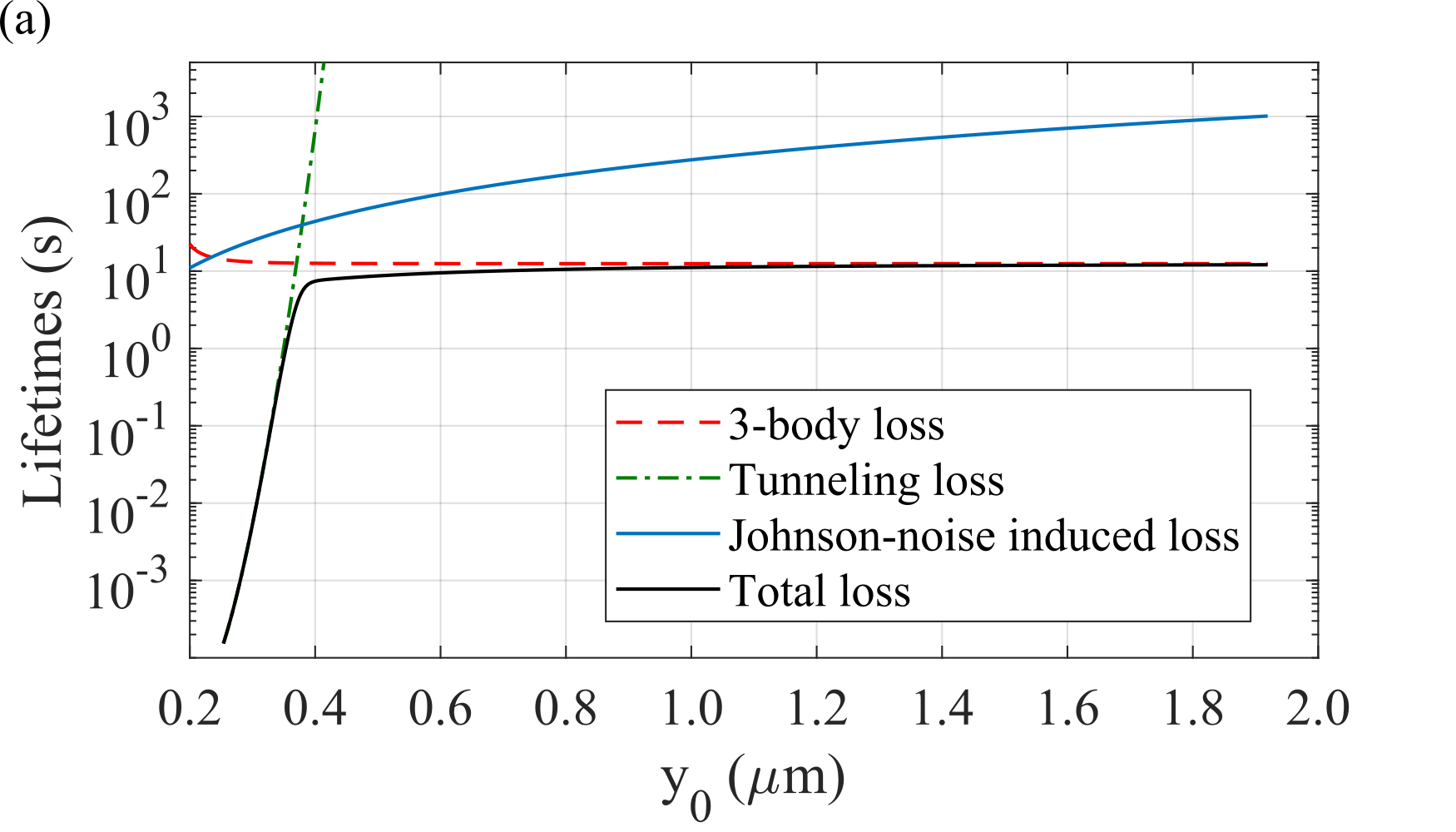}
\includegraphics[width = \linewidth]{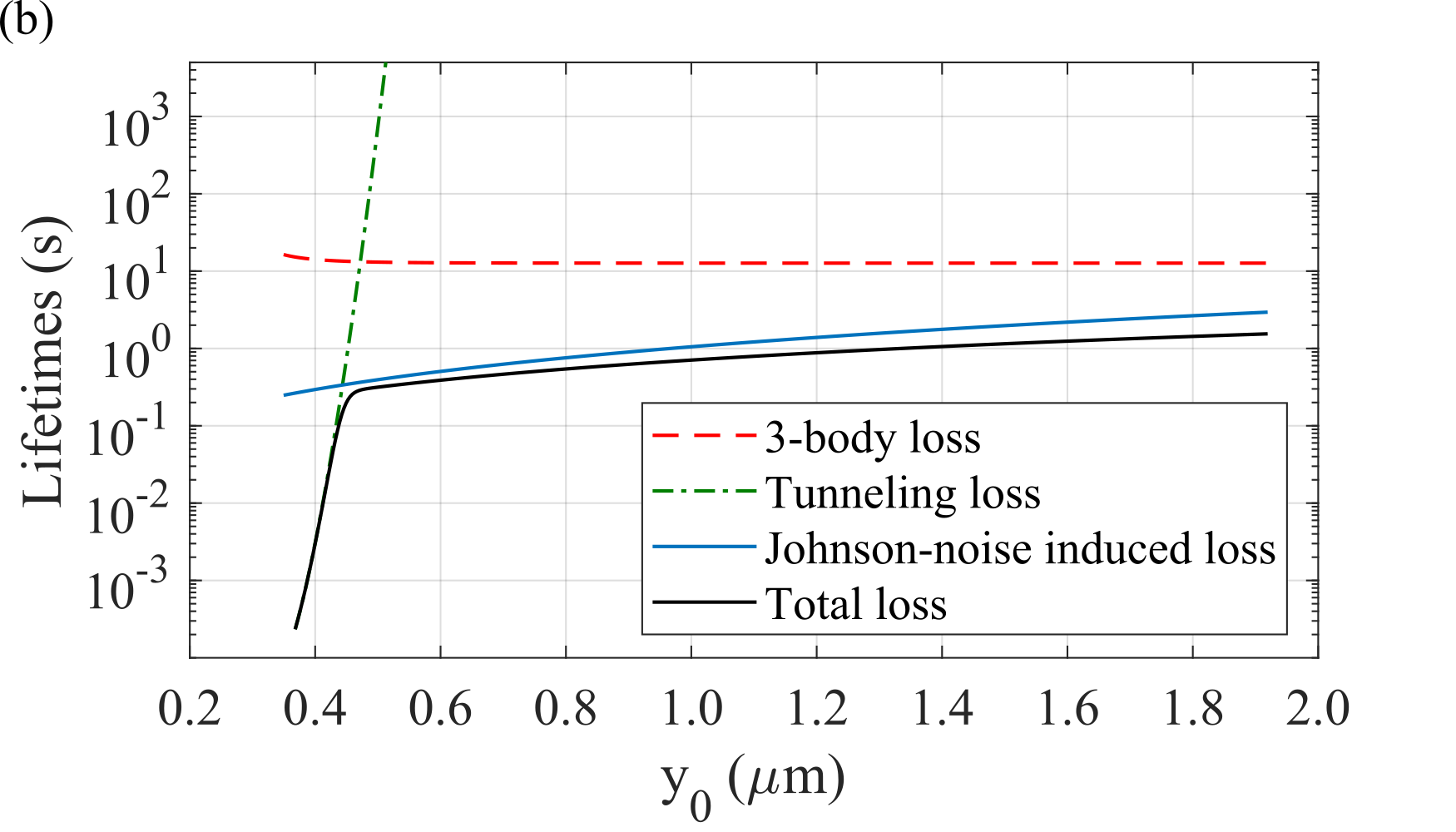}
\caption{Lifetimes calculated (from Eqs. \eqref{eq:Johnson noise induced loss rate} to \eqref{eq:total_life_time_analytic}) versus the position of the harmonic trap center, $y_{0}$, for an \textsuperscript{87}Rb quasi-condensate trapped near (a) a graphene monolayer and (b) a $1$ $\SI{}{\micro\metre}$-thick gold slab, for three different loss mechanisms: 3-body processes (dashed red curves); tunneling losses (dot-dashed green curves); Johnson noise-induced losses (solid blue curves). The total lifetime is shown by the solid black curves. For the graphene monolayer, the total lifetime is limited by 3-body losses to $\sim \SI{12.5}{\second}$, whereas for the gold slab the total lifetime is limited by Johnson noise. Parameter: $\omega_{y} = 2\pi\times\SI{20}{kHz}$.}
\label{fig:Total_lifetimes_Gr_and_Gold}
\end{figure}

Fig.~\ref{fig:Total_lifetimes_Gr_and_Gold} shows the lifetimes resulting from each of the three loss mechanisms considered in the previous section, together with the total lifetime, calculated versus the position of the harmonic trap center, $y_{0}$, for \textsuperscript{87}Rb atoms near (a) a graphene monolayer and (b) the $1$ $\SI{}{\micro\metre}$-thick gold slab. Note that in this figure and henceforth, all the lifetimes are calculated using Eqs. \eqref{eq:Johnson noise induced loss rate} to \eqref{eq:total_life_time_analytic}, which account for the minimum experimentally-detectable atom density, whereas those shown in Figs. \ref{fig:Tunnel_lifetime_main} and \ref{fig:Johnson_Gr_gold} are calculated using Eqs. \eqref{eq:tunnelling loss lifetime}, \eqref{eq:Splin-flip rate} and \eqref{eq:Lifetime} to facilitate comparison with the corresponding lifetimes reported in \cite{Lin2004}. The unperturbed transverse trapping frequency, $\omega_{y} = 2\pi\times\SI{20}{kHz}$, is used in all cases.\\ 

\begin{figure}[ht]
\includegraphics[width = \linewidth]{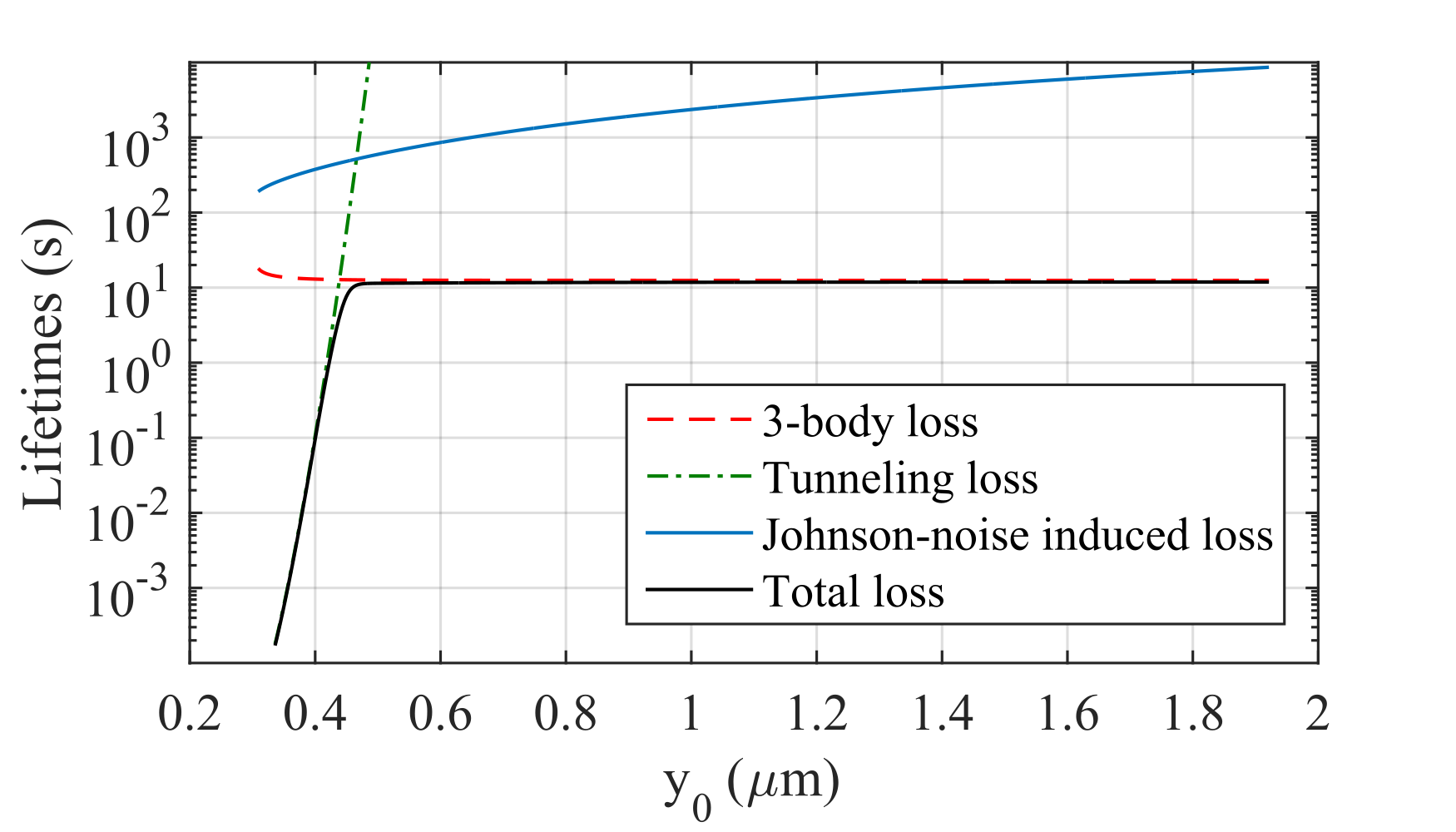}
\caption{Lifetimes calculated (from Eqs. \eqref{eq:Johnson noise induced loss rate} to \eqref{eq:total_life_time_analytic}) versus the position of the harmonic trap center, $y_{0}$, for an \textsuperscript{87}Rb quasi-condensate trapped near an hBN-encased graphene heterostructure for three different loss mechanisms: 3-body processes (dashed red curve); tunneling losses (dot-dashed green curve); Johnson noise-induced losses (solid blue curve). The total lifetime is shown by the solid black curve. For $y_{0} > \SI{0.5}{\micro\metre}$, where the magnetic trap has a well-defined barrier on the side near the surface, the lifetimes are virtually identical to those for a single layer of graphene. Parameters: $\omega_{y} = 2\pi\times\SI{20}{kHz}$, hBN thickness $= \SI{10}{\nano\metre}$.}
\label{fig:Total_lifetimes_hBN_Gr}
\end{figure}

Considering Fig. \ref{fig:Total_lifetimes_hBN_Gr}, we firstly note that the 3-body loss lifetimes (dashed red curves) are identical for the two structures, as expected from Eq. \eqref{eq:1D loss rate_main_text}. Secondly, as a consequence of weaker CP attraction, the tunneling loss lifetime for the graphene monolayer is longer than for the gold slab (compare dot-dashed green curves) and the minimum atom-surface trapping distance is lower. Thirdly, significant improvement in the Johnson noise-limited lifetime is apparent for the graphene monolayer. Whereas Johnson noise in the metal wire limits the total atom lifetime, for graphene the 3-body lifetime of $\sim \SI{12.5}{\second}$ is the limiting factor and Johnson noise is insignificant.

Figure \ref{fig:Total_lifetimes_hBN_Gr} shows lifetimes calculated for \textsuperscript{87}Rb atoms near the hBN-graphene heterostructure, taking the same trapping frequency as in Fig.~\ref{fig:Total_lifetimes_Gr_and_Gold}. The only noticeable difference in the lifetimes compared with those for a graphene monolayer alone relates to tunneling loss (dashed red curve). For given $y_0$, the hBN-graphene structure generates a higher CP potential and, hence, a shorter tunneling lifetime and a higher minimum distance from the trap center to the surface. The Johnson noise is insensitive to the addition of the hBN cladding layers because such layers change neither the number nor mobility of the free charge carriers in the graphene and the total lifetime is still limited by the 3-body loss mechanism. 

\begin{figure}[ht]
\includegraphics[width = \linewidth]{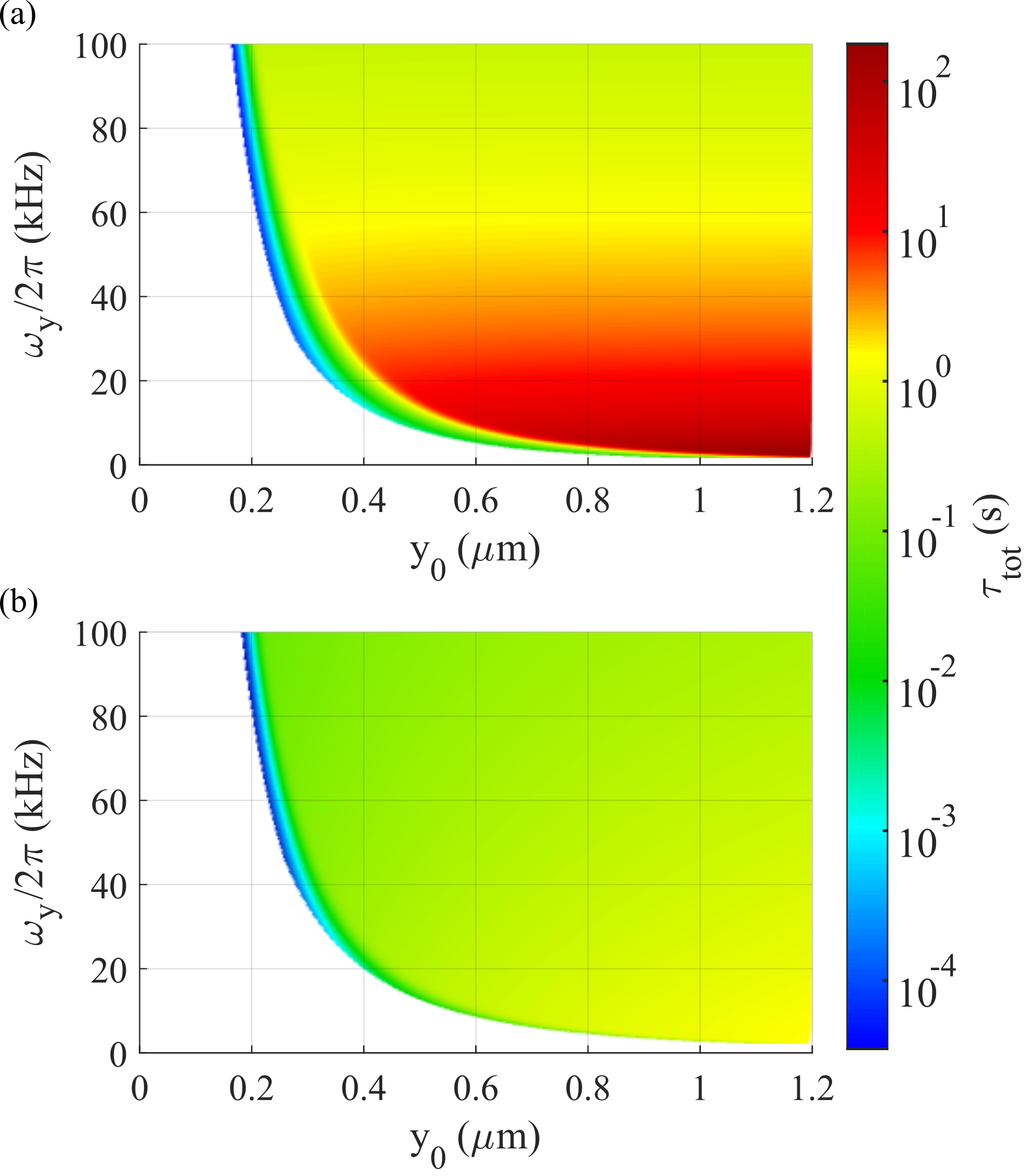}
\caption{Color maps of total lifetimes, calculated (from Eqs. \eqref{eq:Johnson noise induced loss rate} to \eqref{eq:total_life_time_analytic}) versus $y_0$ and $\omega_{y}/2\pi$ for (a) an hBN-encased graphene heterostructure, (b) a $1$ $\SI{}{\micro\metre}$-thick gold slab with a common color scale (right). The lifetime is not defined in the white region because the CP potential distorts the harmonic magnetic potential (i.e. reduces the barrier nearest to the surface) so strongly that the trap cannot be formed. For any given $y_0$ and $\omega_{y}/2\pi$ values, the lifetime for the hBN-graphene structure is longer than for the thin gold slab.}
\label{fig:2D_total_lifetime_GrhBN_Gold}
\end{figure}

Figure \ref{fig:2D_total_lifetime_GrhBN_Gold} shows color maps of the total lifetimes, calculated versus the position of the trap center from the surface and the transverse trapping frequency for (a) the hBN-graphene heterostructure and (b) the $1$ $\SI{}{\micro\metre}$-thick gold slab. The color scale is logarithmic and is common to (a) and (b). Whereas the total lifetime for the thin gold slab is mainly below $\SI{1}{\second}$ (yellow-green in color scale), for the hBN-graphene structure it exceeds $\SI{100}{\second}$ (red shading) for high $y_0$ and low $\omega_y$ values.   

\section{Possible Implementations}

In order to see whether a close surface magnetic trap can be realized using graphene wires we assume a simple side guide configuration, consisting of graphene wire carrying a current, $I$, and a bias magnetic field of magnitude $\abs{\mathbf{B}_{b}}$. This will form a magnetic minimum at a distance

\begin{equation}
y_{0} = \frac{\mu_{0}}{2\pi}\frac{I}{\abs{\mathbf{B}_{b}}}
\label{eq:trap_distance}
\end{equation}
from the graphene sheet. Using the derivation given in \cite{Folman2002} and assuming that the trapping distance is larger then the width of the graphene wire, the trapping frequency is approximated by

\begin{equation} 
\omega_{y} = 2\pi \sqrt{\frac{\mu_{B}g_{F}m_{F}}{m \abs{\mathbf{B}_{0}}}} \frac{\abs{\mathbf{B}_{b}}^2}{I\mu_{0}},
\label{eq:trap_freq}
\end{equation}
where $\abs{\mathbf{B}_{0}}$ is the magnitude of the offset magnetic field parallel to the direction of current flow used to avoid Majorana spin-flips. A trap frequency of $\omega_{y} \approx 2\pi\times\SI{20}{kHz}$ is therefore realizable at a distance of $\SI{400}{\nano\metre}$ with a total current of $\SI{0.7}{\micro A}$ in addition to a bias field of $\SI{35}{\micro T}$ and an offset (Ioffe) field of $\SI{80}{\micro T}$. Since exfoliated and epitaxial graphene on bulk substrates can support current densities in excess of $\sim$ 1000 A/m even in an ultra-high vacuum \cite{Breakdown1,Breakdown2}, and current densities as high as $\sim$ 700 A/m have been reported for free-standing monolayer graphene \cite{Bolotin2}, a graphene conducting channel only 50 nm wide would be sufficient to ensure trap operation with negligible heating.  Wires with larger widths could increase the possible trapping frequencies or enable trapping further from the surface, which may assist with loading the trap. We note, however, that such a trap could not be loaded directly but would need to be mounted on a carrier chip featuring thick metal wires, which generate the field used initially to cool and trap the atoms. This carrier chip must be placed far enough from the graphene and the atoms to produce negligible Johnson noise and CP attraction effects, but also close enough to create a sufficiently compressed trap. Given a $\SI{50}{\micro\metre}$ separation between the atom cloud and the carrier chip, gold wires carrying a current of 1A could produce the transverse trap frequency of $\omega_{r} = 2\pi\times\SI{20}{kHz}$ needed for the traps shown in Figs. 4, 6, 9 and 10. Since thin van der Waals heterostructures are almost transparent, laser light can pass through them and be retro-reflected from a metal coating on the carrier chip in order to form a mirror MOT.\\

In an alternative configuration, the potential for trapping the atoms could be provided by optical fields, for example an electromagnetic standing wave generated by on-chip mounted optics. In this case, the graphene wires could be used to perturb strongly the optical potential or act as a source of magnetic fields to enable, for example, tuning the scattering length via Feshbach resonances.\\

Another issue that has been observed when trapping atoms close to a surface, especially metal, is the effect of stray electric fields originating from the polarization of adsorbed atoms \cite{Hunger_Camerer_2010}. Although this effect has not yet been measured for graphene surfaces, covering the graphene layer with a dielectric layer such as hBN is expected to suppress these effects by limiting polarization of any adsorbates and keeping them away from the graphene layer(s), so preventing them doping it.\\

Graphene-based atoms chips could be fabricated by MBE growth of graphene \cite{MBEgrowth1,MBEgrowth2,MBEgrowth3}, or deposition of exfoliated graphene on hBN, followed by selective etching of the graphene to define the conducting channel and, finally, deposition of capping layers of hBN either by epitaxial growth or by placing exfoliated hBN layers, as now widely done to create van der Waals heterostructures \cite{vdW1,vdW2}.\\


\section{Conclusions}
In summary, we have presented a general formalism for calculating how the lifetime of an atomic quantum gas is affected by the Johnson noise and atom-surface CP attraction of van der Waals heterostructures comprising arbitrary configurations of 2D materials such as graphene. The electromagnetic reflection coefficients and corresponding electrical conductivities of the 2D layers are of crucial importance in determining both the Johnson noise and CP potential. Since both of these parameters are lower for graphene than for the metal layers generally used in atom-chips, so too are the Johnson noise and CP atom-surface attraction. Consequently, for given atom-surface separation, the spin-flip and tunneling loss rates are both lower near graphene-based van der Waals heterostructures than near metal wires, meaning that such heterostructures can improve the performance of atom-chips. For example, although high Johnson noise limits the lifetime of atom clouds trapped between 0.4 and $\SI{2}{\micro\metre}$ from the chip surface to less than 1s, such noise is negligible for atoms trapped near graphene, whose lifetime can, in principle, reach $\sim$ 100 s, limited only by 3-body losses and background gas collisions. For atom-surface separations below $\SI{0.4}{\micro\metre}$, the lifetime of the atom cloud is limited by tunneling losses for both metallic and 2D conductors. However, due to the weak CP atom-surface attraction, such losses are lower near van der Waals heterostructures; around 4 orders of magnitude lower for atoms held $\SI{0.4}{\micro\metre}$ from the surface.\\

As a result of their favourable noise and CP characteristics, van der Waals heterostructures offer a solid-state solution to the long-standing challenge of holding ultracold atom clouds closer than $\SI{1}{\micro\metre}$ from an atom-chip surface for long enough (up to 100 s) to perform various experiments and measurements on the atom cloud. Moreover, van der Waals heterostructures that enable robust sub-$\SI{}{\micro\metre}$ atom trapping would control atomic condensates on length scales that are smaller than presently achievable optically and below the healing length, thereby providing access to new regimes of quantum many-body physics. The ability to achieve long lifetimes for ultracold atoms held as close as 400 nm to an electronic device also opens a route to creating new hybrid atomic-solid state quantum systems, for example a Rydberg atom coupled to a quantum dot formed within a 2D conductor \cite{Mancsdot,Mancsdot2}. Since the micron-scale confinement length of electrons in the quantum dot is similar to that of the excited electron in the Rydberg atom, new types of electron orbital, shared between the atom and the condensed matter parts of the system may be created. Such hybrid states may yield new regimes of quantum control and information storage/processing, for example relating to side-band cooling of graphene \cite{Miskeen_Khan}.



\appendix
\section{Dyadic Green's function}
\label{supplement: Green's function}
In this Appendix, we discuss Green's tensors for planar multilayer systems, like those considered in the main text. We will start by considering the general characteristics of an electric field in a homogeneous space, in order to provide an intuitive explanation of the Green's tensors.\\

\subsection{General definition}
Let us recall the inhomogeneous Helmholtz equation describing the relationship between an electric field, $\mathbf{E}$, and a current density, $\mathbf{j}$, of angular frequency $\omega$ at position $\mathbf{r}$ with respect to a linear, isotropic, inhomogeneous medium, with relative magnetic permeability, $\mu(\omega)$, and relative permittivity $\epsilon(\omega)$ \cite{crosse_fuchs_buhmann_2015}:
\begin{equation}
\curl{\curl{\mathbf{E}(\omega, \mathbf{r})}} - k^{2}\mathbf{E}(\omega, \mathbf{r}) = \mathrm{i}\omega\mu_{0}\mu\mathbf{j}(\omega, \mathbf{r}).
\label{eq:wave_eq}
\end{equation}
Here, $k = \sqrt{\epsilon(\omega)\mu(\omega)\omega^{2}/c^{2}}$ is the magnitude of the wave vector associated with the electromagnetic wave, $\mu_{0}$ is the permeability of free space, and  
\begin{equation}
\mathbf{j}(\omega, \mathbf{r}) = j_{x}(\omega, \mathbf{r})\mathbf{e}_{x} + j_{y}(\omega, \mathbf{r})\mathbf{e}_{y} + j_{z}(\omega, \mathbf{r})\mathbf{e}_{z},
\label{eq:current_density}
\end{equation}
where $\mathbf{e}_{x}, \mathbf{e}_{y}$ and $\mathbf{e}_{z}$ are the unit vectors in the $x$-, $y$- and $z$-direction, respectively, with $j_{x}$, $j_{y}$ and $j_{z}$ being the corresponding current density components.\\

\begin{figure}[ht]
\centering
\includegraphics[width = \linewidth]{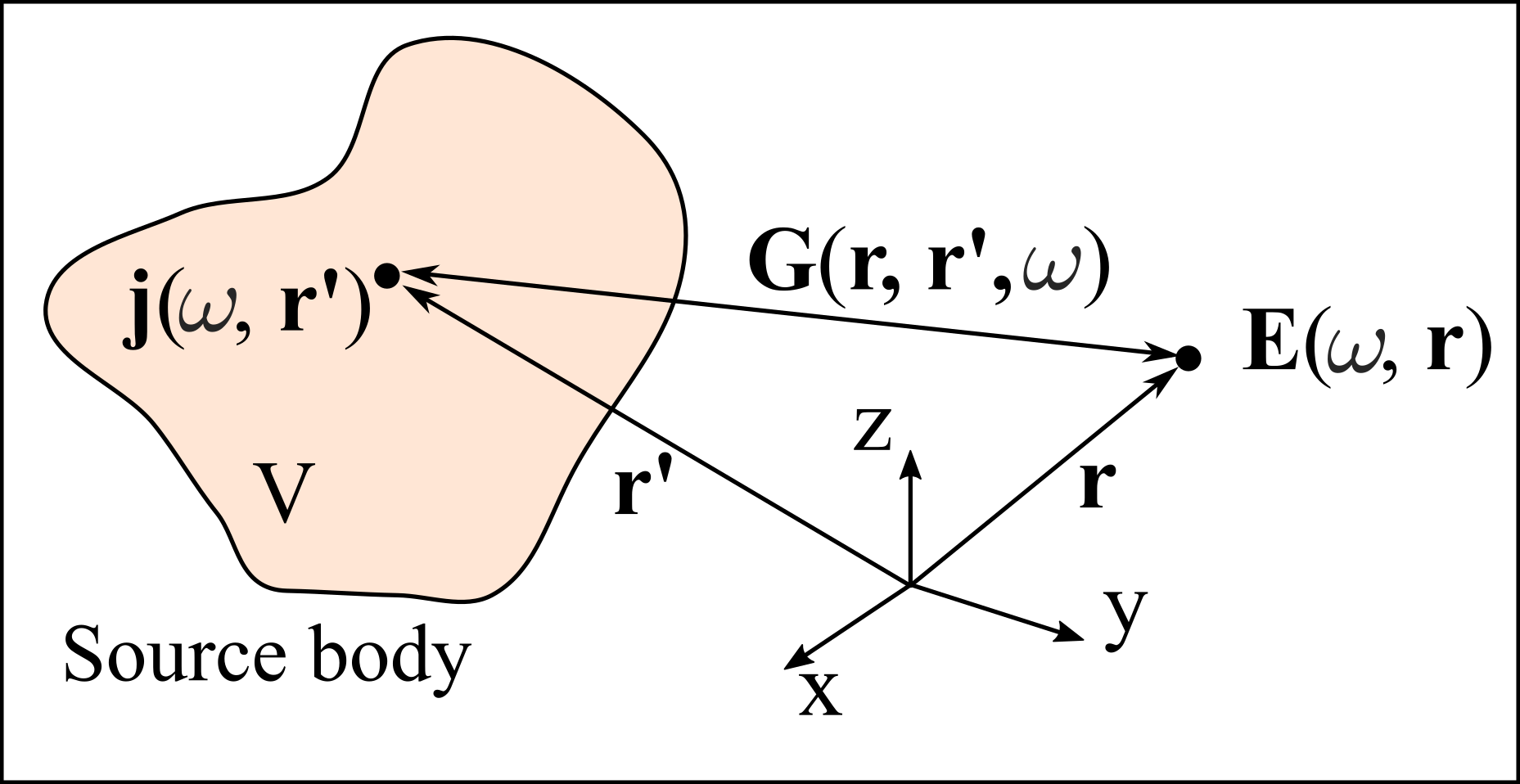}
\caption{Illustration of the dyadic Green's function $\mathbf{G}(\mathbf{r}, \mathbf{r}^{\prime}, \omega)$. The Green's function relates the local current source, $\mathbf{j}$, at point, $\mathbf{r}^{\prime}$, and the associated electric field $\mathbf{E}$ at point $\mathbf{r}$. The total electric field is the superposition of every field corresponding to each point source in the source body of volume $V$.}
\label{fig:Souce_field}
\end{figure}

The classical Green's tensor $\mathbf{G} = (\mathbf{G}_{x}, \mathbf{G}_{y}, \mathbf{G}_{z})$, as a function of the field point position, $\mathbf{r}$, the source point position, $\mathbf{r}^{\prime}$, and the wave angular frequency, $\omega$, is the unique solution to the following differential equations:
\begin{equation}
\curl{\curl{\mathbf{G}_{x}(\mathbf{r}, \mathbf{r}^{\prime}, \omega)}} - k^{2}\mathbf{G}_{x}(\mathbf{r}, \mathbf{r}^{\prime}, \omega) = \delta(\mathbf{r} - \mathbf{r}^{\prime})\mathbf{e}_{x}, 
\label{eq:G_x}
\end{equation}
\begin{equation}
\curl{\curl{\mathbf{G}_{y}(\mathbf{r}, \mathbf{r}^{\prime}, \omega)}} - k^{2}\mathbf{G}_{y}(\mathbf{r}, \mathbf{r}^{\prime}, \omega) = \delta(\mathbf{r} - \mathbf{r}^{\prime})\mathbf{e}_{y},
\label{eq:G_y}
\end{equation}
\begin{equation}
\curl{\curl{\mathbf{G}_{z}(\mathbf{r}, \mathbf{r}^{\prime}, \omega)}} - k^{2}\mathbf{G}_{z}(\mathbf{r}, \mathbf{r}^{\prime}, \omega) = \delta(\mathbf{r} - \mathbf{r}^{\prime})\mathbf{e}_{z},
\label{eq:G_z}
\end{equation}
where $\delta(\mathbf{r} - \mathbf{r}^{\prime})$ is the Dirac delta function. These three Green's functions, in a column vector form, can be combined into a single tensor, giving the following general definition of the dyadic Green's function (Green's tensor) for the electric field:  
\begin{equation}
\curl{\curl{\mathbf{G}(\mathbf{r}, \mathbf{r}^{\prime}, \omega)}} - k^{2}\mathbf{G}(\mathbf{r}, \mathbf{r}^{\prime}, \omega) = \mathbf{I}\delta(\mathbf{r} - \mathbf{r}^{\prime}),
\label{eq:general_G}
\end{equation}
where $\mathbf{I}$ is the unit dyad (unit tensor).\\

We can see that each column of the tensor $\mathbf{G}$ can be mathematically treated individually: the curl operators can be applied to any column of the Green's tensor as if they were to act on a single column vector. In addition, each column can be interpreted separately: the first column of the Green's tensor describes the field due to a point source in the $x$-direction, the second column the field due to a point source in the $y$-direction, and the third column the field due to a point source in the $z$-direction.\\

Consequently, a particular solution of Eq. \eqref{eq:wave_eq} defined by the dyadic Green's function is 
\begin{equation}
\mathbf{E}(\omega, \mathbf{r}) = \mathrm{i}\omega\mu\mu_{0}\int_{V}\mathbf{G}(\mathbf{r}, \mathbf{r}^{\prime}, \omega) \mathbf{j}(\omega, \mathbf{r}^{\prime})\mathrm{d}^{3}r^{\prime},
\label{eq:particular_E}
\end{equation}\\
where the integral is evaluated over the volume $V$ of the current source body (see Fig.~\ref{fig:Souce_field}).

\subsection{Green's tensor for planar multilayer systems}
Let us consider the situation shown in Fig.~\ref{fig:dipole_interface}, where a radiating electric dipole is located above a layered substrate. We assume that the upper half-space is vacuum while the lower half-space (atom-chip substrate) is optically denser.\\

Since the electric field in layer 1 is the superposition of the field directly radiated from the dipole and the field scattered by the material layers, the Green's function can, correspondingly, be decomposed into two contributions: a Green's function for homogeneous space and a scattering Green's function reflecting dielectric inhomogeneity. In order to find the primary dyadic Green's function, $\mathbf{G}^{(0)}(\mathbf{r}, \mathbf{r}^{\prime}, \omega)$, (sometimes called the ``free-space Green's function"), we remove the interfaces in Fig.~\ref{fig:dipole_interface} and assume that the electric dipole, $\mathbf{d}$, located at $\mathbf{r}^{\prime} = (x^{\prime}, y^{\prime}, z^{\prime})$, is in a homogeneous, linear and isotropic medium, characterized by permittivity and permeability functions $\epsilon_{1}(\omega)$ and $\mu_{1}(\omega)$. The superscript $(0)$ here is to remind us that this Green's function is the primary Green's function. The associated electric field at $\mathbf{r} = (x, y, z)$, and its corresponding wave vectors, respectively, are 
\begin{equation}
\mathbf{E}(\mathbf{r}, \omega) = \omega^{2}\mu_{0}\mu_{1}\mathbf{G}^{(0)}(\mathbf{r}, \mathbf{r}^{\prime}, \omega)\mathbf{d}(\mathbf{r}^{\prime}, \omega),
\label{eq:E_and_G0}
\end{equation}
\begin{equation}
\mathbf{k}_{1}(\omega) = k_{x}\mathbf{e}_{x} + k_{y1}\mathbf{e}_{y} + k_{z}\mathbf{e}_{z} =  k^{\parallel}\mathbf{e}_{k^{\parallel}} + k_{y1}\mathbf{e}_{y},
\label{eq:wave_vector1}
\end{equation}
where $k_{x}$ and $k_{z}$ are the $x$- and $z$-components of the wavevector $k^{\parallel}\mathbf{e}_{k^{\parallel}} = k_{x}\mathbf{e}_{x} + k_{z}\mathbf{e}_{z}$ in the $x-z$-plane, which are the same in every layer, and $k_{y1}$ is the $y$-component of the wave vector in layer 1.\\

\begin{figure}[ht]
\centering
\includegraphics[width = \linewidth]{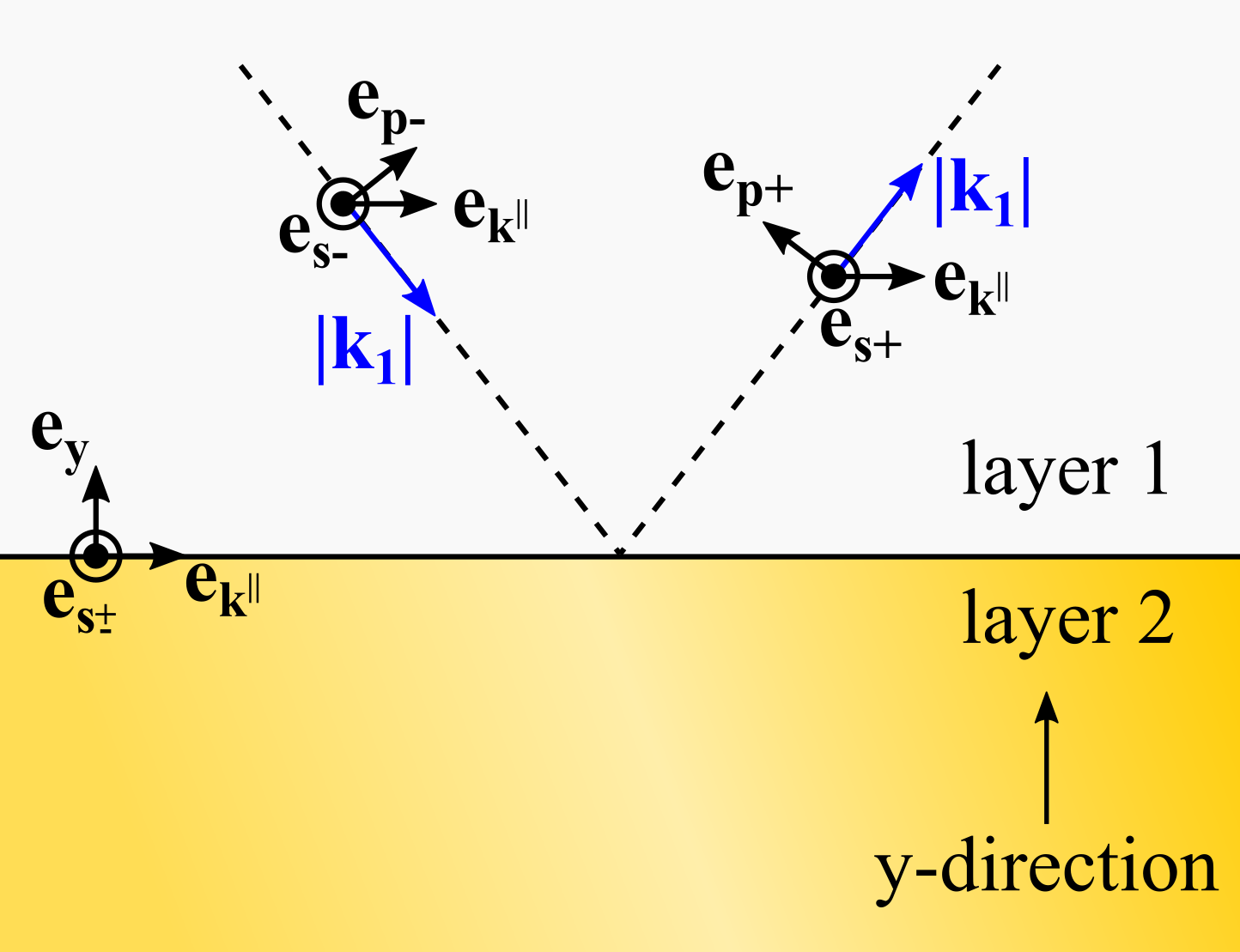}
\caption{Definition of the polarization unit vectors. The plane spanned by the vector $\mathbf{e}_{y}$, pointing in the positive $y$-direction, and the vector $\mathbf{e}_{k^{\parallel}} = (k_{x}, 0, k_{z})/k^{\parallel}$ defines the plane of incidence. The $s$-polarization unit vectors, $\mathbf{e}_{s\pm}$, are perpendicular to the plane of incidence, whilst the $p$-polarization unit vectors, $\mathbf{e}_{p\pm}$, are parallel to the plane of incidence and perpendicular to the directions of propagation (blue arrows).}
\label{fig:definition_of_p_s_pol}
\end{figure}

The primary Green's tensor can be written in the form \cite{Novotny2006, Goncalves2016, Buhmann_ii}
\begin{multline}
\mathbf{G}^{(0)}(\mathbf{r}, \mathbf{r}^{\prime}, \omega) = 
\frac{\mathrm{i}}{8\pi^{2}}\iint_{-\infty}^{\infty}\frac{1}{k_{y1}}
\mathbf{M}^{(0)}(k_{x}, k_{z})\times\\
\mathrm{e}^{\mathrm{i}[k_{x}(x - x^{\prime}) + k_{z}(z - z^{\prime}) + k_{y1}\abs{y - y^{\prime}}]}\mathrm{d}k_{x}\mathrm{d}k_{z},
\label{eq:G0_implicit}
\end{multline}
in which
\begin{equation}
\mathbf{M}^{(0)}(k_{x}, k_{z}) = (\mathbf{e}_{s\pm}\otimes\mathbf{e}_{s\pm}) + (\mathbf{e}_{p\pm}\otimes\mathbf{e}_{p\pm}),
\label{eq:M_implicit}
\end{equation}
where $\otimes$ represents a tensor product.
Here, the polarization unit vectors for $s$- and $p$-polarized waves in layer 1 are defined as (see Fig.~\ref{fig:definition_of_p_s_pol})
\begin{align}
\mathbf{e}_{s\pm} &= \mathbf{e}_{k^{\parallel}}\times \mathbf{e}_{y},\\
\mathbf{e}_{p\pm} &= \frac{1}{k_{1}}(k^{\parallel}\mathbf{e}_{y}\mp k_{y1}\mathbf{e}_{k^{\parallel}}), 
\label{eq:pol_unit_vectors}
\end{align}
where $k_{1}  = \sqrt{\epsilon_{1}(\omega)\mu_{1}}(\omega)\omega/c = (k_{y1}^{2} + k^{\parallel2})^{1/2}$ is the wave number and the upper (-) sign applies for $y > y^{\prime}$ (i.e. waves propagating in the positive $y$-direction), whilst the lower (+) sign applies for $y < y^{\prime}$ (i.e. waves propagating in the negative $y$-direction).

Let us now consider what happens to the electric field described by the above primary Green's function when a reflective planar layered structure is added to the system below the dipole, as depicted in Fig.~\ref{fig:dipole_interface}. Physically, the role of the layered structure is to reflect back the electromagnetic wave radiated from the dipole. Therefore, mathematically, multiplying the individual incident plane waves in $\mathbf{G}^{(0)}$ with the corresponding generalized Fresnel reflection coefficients, $r_{s}^{(1)}$ and $r_{p}^{(1)}$, along with changing the phase factor (the exponential term) accordingly, yields the scattering Green's tensor: 
\begin{multline}
\mathbf{G}^{(1)}(\mathbf{r}, \mathbf{r}^{\prime}, \omega) = 
\frac{\mathrm{i}}{8\pi^{2}}\iint_{-\infty}^{\infty}\frac{1}{k_{y1}}
\mathbf{M}^{(1)}(k_{x}, k_{z})\\
\times \mathrm{e}^{\mathrm{i}[k_{x}(x - x^{\prime}) + k_{z}(z - z^{\prime}) + k_{y1}(y + y^{\prime})]}\mathrm{d}k_{x}\mathrm{d}k_{z},
\label{eq:Green_reflected_elec}
\end{multline}
where 
\begin{equation}
\mathbf{M}^{(1)}(k_{x}, k_{z}) = r_{s}^{(1)}(\mathbf{e}_{s+}\otimes\mathbf{e}_{s-}) + r_{p}^{(1)}(\mathbf{e}_{p+}\otimes\mathbf{e}_{p-}).
\label{eq:M_implicit_1}
\end{equation}
Here, the superscript $(1)$ is to remind us that the Green's tensor is the scattering Green's tensor for layer 1.\\

The total electric field due to a radiating \emph{electric} dipole above a planar structure can now be written as

\begin{equation}
\mathbf{E}(\mathbf{r}, \omega) = \omega^{2}\mu_{0}\mu_{1}\mathbf{G}_{E}(\mathbf{r}, \mathbf{r}^{\prime}, \omega)\mathbf{d}(\mathbf{r}^{\prime}, \omega),
\label{eq:total_E_field}
\end{equation}
where $\mathbf{G}_{E}(\mathbf{r}, \mathbf{r}^{\prime}, \omega) = \mathbf{G}^{(0)}(\mathbf{r}, \mathbf{r}^{\prime}, \omega)+\mathbf{G}^{(1)}(\mathbf{r}, \mathbf{r}^{\prime}, \omega)$ and we introduce a subscript $E$ to emphasize that this Green's function is for an electric dipole.\\

In order to obtain a scattering Green's tensor for a \emph{magnetic} dipole, we interchange the Fresnel reflection coefficients in \eqref{eq:M_implicit}.\\


The relevant Green's tensors for calculating the CP potential and the Johnson noise must be evaluated at $\mathbf{r} = \mathbf{r}^{\prime} = \mathbf{r}_{0}$, where $\mathbf{r}_{0} = (x_{0}, y_{0}, z_{0})$ is the position of the center of the magnetic trap. After straightforward manipulation of the polar coordinates, the equal-position scattering Green's tensor is, therefore, given by \cite{Buhmann_ii, amorim_2017}
\begin{multline}
\mathbf{G}^{(1)}(\mathbf{r}_{0}, \mathbf{r}_{0}, \omega) = \frac{\mathrm{i}}{8\pi}\int_{0}^{\infty}\mathrm{d}k^{\parallel}\frac{k^{\parallel}}{k_{y1}}\mathrm{e}^{2\mathrm{i}k_{y1}y_{0}}\\
\times \Big[\mathbf{M}_{\alpha}r_{s}^{(1)}(k^{\parallel}, \omega)
+ \frac{c^{2}}{\omega^{2}}
\mathbf{M}_{\beta}r_{p}^{(1)}(k^{\parallel}, \omega)\Big],
\label{eq:Green_for_electric_dipole}
\end{multline}
where $y_{0}$ is the shortest distance between the surface and the center of the magnetic trap and
\begin{equation}
k_{y1} = \Big(\mu_{1}\epsilon_{1}\frac{\omega^{2}}{c^{2}} - k^{\parallel2}\Big)^{1/2},
\label{eq:k_y1}
\end{equation}
with $k^{\parallel2} = k_{x}^{2} + k_{z}^{2}$. 
The tensors $\mathbf{M}_{\alpha}$ and $\mathbf{M}_{\beta}$ in Eq. \eqref{eq:Green_for_electric_dipole} are given by

\begin{align}
\mathbf{M}_{\alpha} = & 
\begin{pmatrix}
1 & 0 & 0\\
0 & 0 & 0\\
0 & 0 & 1
\end{pmatrix},\label{eq:M_alpha}\\
\mathbf{M}_{\beta} = & 
\begin{pmatrix}
-k_{y1}^{2} & 0 & 0\\
0 & 2k^{\parallel2} & 0\\
0 & 0 & -k_{y1}^{2}
\end{pmatrix}.
\label{eq:M_beta}
\end{align}
Note that the forms of $\mathbf{M}_{\alpha}$ and $\mathbf{M}_{\beta}$ depend on the coordinate system used and that Equation \eqref{eq:Green_for_electric_dipole} only describes an electromagnetic field with a \emph{real} frequency, created by a radiating \emph{electric} dipole.\\

For a purely imaginary frequency, $\omega = \mathrm{i}\xi$, where $\xi$ is real, e.g. the Matsubara frequencies that appear in the CP potential calculations, the wave vector in the direction perpendicular to the surface is always purely imaginary, $k_{y1} = \mathrm{i}\kappa_{1}^{\perp}$, with
\begin{equation}
\kappa_{1}^{\perp}(\mu_{1}, \epsilon_{1}, \mathrm{i}\xi) = \sqrt{\mu_{1}(\mathrm{i}\xi)\epsilon_{1}(\mathrm{i}\xi)\frac{\xi^{2}}{c^{2}} + k^{\parallel2},}
\label{eq:kappa_1}
\end{equation}
and 
\begin{equation}
k^{\parallel} = \sqrt{\kappa_{1}^{\perp2} - \mu_{1}\epsilon_{1}\frac{\xi^{2}}{c^{2}}} = \sqrt{\kappa_{l}^{\perp2} - \mu_{l}\epsilon_{l}\frac{\xi^{2}}{c^{2}}},
\label{eq:k_ll}
\end{equation}
where the subscript $l$ denotes the layer index corresponding to each wave vector. Equations \eqref{eq:kappa_1} and \eqref{eq:k_ll} tell us that the wave numbers in the direction perpendicular to the surface are functions of the optical properties of the materials and the incident wave frequencies, whereas the wave numbers in the direction parallel to the surface are constant for a given $\kappa_{l}^{\perp}$.    
The equal-position Green's tensor for \emph{purely imaginary} frequencies is then given by \cite{Novotny2006, Buhmann_ii, crosse_fuchs_buhmann_2015} 
\begin{multline}
\mathbf{G}^{(1)}(\mathbf{r}_{0}, \mathbf{r}_{0}, \mathrm{i}\xi) = \frac{1}{8\pi}\int_{\xi/c}^{\infty}\mathrm{d}\kappa_{1}^{\perp}\mathrm{e}^{-2\kappa_{1}^{\perp}y_{0}} \\ 
\times \Big[\mathbf{M}_{\alpha}r_{s}^{(1)}(k^{\parallel}, \mathrm{i}\xi)
- \frac{c^{2}}{\xi^{2}}
\mathbf{M}_{\beta}r_{p}^{(1)}(k^{\parallel}, \mathrm{i}\xi)\Big].
\label{eq:Green_for_CP}
\end{multline}

Finally, we note that the Green's tensor for a magnetic dipole, which is needed for the Johnson noise calculations, can be readily obtained by interchanging the reflection coefficients in \eqref{eq:Green_for_electric_dipole}.

\section{Optical properties of materials}
\label{supplementary: Optical properties}
In this section, we briefly introduce the optical properties
of graphene, hBN and bulk gold. They are used in order to determine the behavior of the reflection coefficients of our system (see Appendix \ref{app:Reflection_coefficients}).\\

Graphene's optical conductivity can be split into two distinct contributions. The first describes how the charge carriers respond to electromagnetic radiation by transitioning to higher energy states within the same energy bands without conserving momentum (intraband-transitions, $\sigma_{\mathrm{intra}}(\omega)$). The second accounts for vertical momentum-conserving transitions from the valence band to the conduction band, induced by the electromagnetic radiation (interband- transitions, $\sigma_{\mathrm{inter}}(\omega)$), where $\omega$ is the angular frequency of the electromagnetic field to which the graphene is exposed.\\

The expression for graphene's conductivity has been considered using multiple approaches and limits (see for example \cite{Falkovsky08,Fialkovsky11}) and the choice of a specific description depends on the features of the system under analysis and/or on the particular aspect under investigation (see for example \cite{Egerland19,Werra16a,Henkel18,Biehs14} and the references below).
For our systems and its parameters we can use the expression derived from the Kubo formula \cite{Goncalves2016, stauber_peres_geim_2008, hanson_2013}, which gives:

\begin{equation}
\sigma_{\mathrm{intra}}(\omega) = \frac{\sigma_{0}}{\pi}\frac{4}{\hbar\gamma -\mathrm{i}\hbar\omega} \bigg[E_{F} + 2k_{B}T \ln \bigg( 1 + e^{-E_{F}/k_{B}T} \bigg) \bigg], 
\label{eq:conductivity_intra}
\end{equation}

\begin{equation}
\sigma_{\mathrm{inter}}(\omega) = \sigma_{0}\bigg[G(\hbar\omega/2) + \mathrm{i}\frac{4\hbar\omega}{\pi}\int_{0}^{\infty}dE\frac{G(E) - G(\hbar\omega/2)}{(\hbar\omega)^{2} - 4E^{2}}\bigg],
\label{eq:conductivity_inter}
\end{equation}
in which

\begin{equation}
G(X) = \frac{\sinh\bigg(\frac{\displaystyle X}{\displaystyle k_{B}T}\bigg)}{\cosh\bigg(\frac{\displaystyle E_{F}}{\displaystyle k_{B}T}\bigg) + \cosh\bigg(\frac{\displaystyle X}{\displaystyle k_{B}T}\bigg)},
\label{eq:G(x)_function}
\end{equation}
where $\sigma_{0} = e^{2}/(4\hbar)$ is the universal alternating-current conductivity of graphene, $\gamma$ is the electron relaxation rate in graphene, $E_{F}$ is the Fermi energy and $T$ is the temperature of the graphene layer.\\

Within the four-parameter semi-quantum model \cite{Musa2018, Lee2014,  Woessner2014, Gervais1974}, the in-plane optical conductivity of an ultra-thin hBN slab comprising a few monolayers is 

\begin{equation}
\sigma_{\mathrm{hBN}}(\mathrm{i}\xi_{j}) = \mathrm{i}\epsilon_{0}\xi_{j} t_{\mathrm{hBN}}[\epsilon_{z, \mathrm{hBN}}(\mathrm{i}\xi_{j}) - \epsilon_{z}(\infty)].
\label{eq:hBN_conductivity}
\end{equation}
Here, the frequency-dependent permittivity of hBN is given by
\begin{equation}
\epsilon_{f, \mathrm{hBN}}(\mathrm{i}\xi_{j}) = \epsilon_{f}(\infty) + \frac{s_{\nu,f}\omega_{\nu,f}^{2}}{\omega_{\nu,f}^{2} + \gamma_{\nu,f}\xi_{j} + \xi_{j}^{2}},
\label{eq:hBN_permit}
\end{equation}
where $f = x, y, z$, $\omega_{\nu,z} = \SI{2.58e14}{rad\per\second}$, $\gamma_{\nu,z} = \SI{1.319e12}{rad\per\second}$, $s_{\nu,z} = 1.83$, $\epsilon_{z}(\infty) = 4.87$ (see the supplementary information of \cite{Cai2007}) and $t_{\mathrm{hBN}}$ is the thickness of the hBN slab.\\ 

For metals, in particular gold, we use the Drude model for the permittivity

\begin{equation}
\epsilon_{\mathrm{metal}}(\mathrm{i}\xi_{j}) = 1 + \frac{\omega_{\mathrm{p}}^{2}}{\xi_{j}^{2} + \Gamma_{D}\xi_{j}},
\label{eq:Drude_model}
\end{equation}
where, for gold, $\omega_{\mathrm{p}} = \SI{1.38e16}{rad\per\second}$ is the plasma frequency and $\Gamma_{D} = \SI{1.075e14}{rad\per\second}$ is the electron relaxation rate \cite{Okamoto2001, Novotny2006}.

\section{Reflection coefficients}
\label{app:Reflection_coefficients}
Typical atom-chips can be modelled as planar multilayer structures (see Fig.~\ref{fig:dipole_interface}) with generalized Fresnel reflection coefficients given by the following recursive relations \cite{Buhmann_ii,Zhang05}

\begin{equation} \label{eq:r_s_multilayer}
\begin{split}
r_{s}^{(l)} & = r_{s}^{(l)}(k^{\parallel}, \omega = \mathrm{i}\xi_{j}) \\
 & = \frac{a+\{b\cdot\mathrm{exp}(2\mathrm{i}k_{yl+1}^{\perp}t_{l+1})r_{s}^{(l+1)}\}}{b+\{a\cdot\mathrm{exp}(2\mathrm{i}k_{yl+1}^{\perp}t_{l+1})r_{s}^{(l+1)}\}},
\end{split}
\end{equation}

\begin{equation} \label{eq:r_p_multilayer}
\begin{split}
r_{p}^{(l)} & = r_{p}^{(l)}(k^{\parallel}, \omega = \mathrm{i}\xi_{j}) \\
 & = \frac{c+\{d\cdot\mathrm{exp}(2\mathrm{i}k_{yl+1}^{\perp}t_{l+1})r_{p}^{(l+1)}\}}{d+\{c\cdot\mathrm{exp}(2\mathrm{i}k_{yl+1}^{\perp}t_{l+1})r_{p}^{(l+1)}\}},
\end{split}
\end{equation}
where $a = (\mu_{l+1}k_{yl}-\mu_{l}k_{yl+1})$, $b = (\mu_{l+1}k_{yl}+\mu_{l}k_{yl+1})$, 
$c = (\epsilon_{l+1}k_{yl}-\epsilon_{l}k_{yl+1})$, and
$d = (\epsilon_{l+1}k_{yl}+\epsilon_{l}k_{yl+1})$ for $l = 1,...,n-1$ with $\mu_{l} = \mu_{l}(\mathrm{i}\xi), \epsilon_{l} = \epsilon_{l}(\mathrm{i}\xi)$ and a termination condition $r_{s,p}^{n} = 0$. Here, $k_{yl}$ is defined in the same manner as Eqs. \eqref{eq:kappa_1} and \eqref{eq:k_ll}. Note that the superscripts, $(l)$, on $r_{s}$ and $r_{p}$ are indices denoting which layers the reflection coefficients correspond to. We use this method to calculate the reflection coefficients of the metallic conducting wires.\\

Another way to determine the reflection coefficients is to use a transfer-matrix method (see, for example, \cite{Zhan2013}). We use this as a convenient method to calculate the reflection coefficients of the structures that incorporate graphene layers. We now present a concise description of this method. There are two basic elements in this formalism, namely, transmission matrices and propagation matrices: a transmission matrix describes the change of the wave amplitudes when the wave crosses an interface between two media, whereas a propagation matrix captures the phase change when the wave propagates through a medium. Important physical quantities in this method are the conductivities of the ultra-thin layers and the permittivities of thick media.\\  

\begin{figure}
\center
\includegraphics{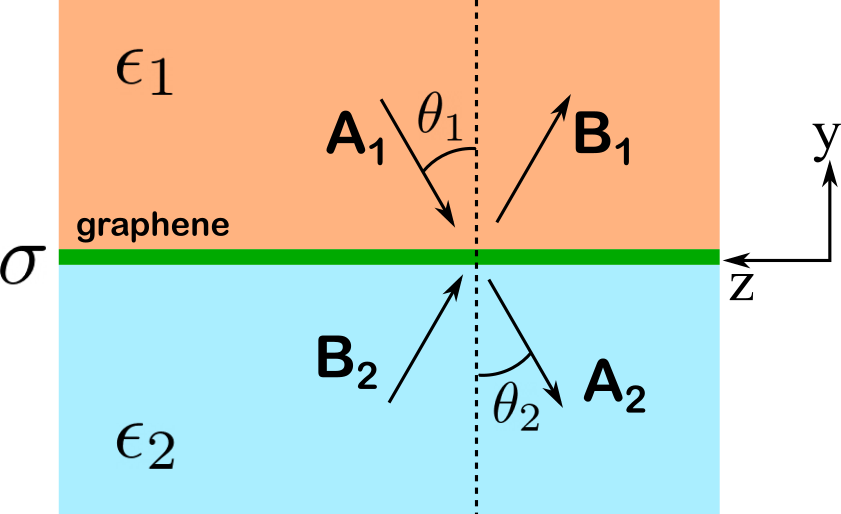}
\caption{A schematic diagram of electromagnetic scattering in a structure composed of a single graphene sheet (green) sandwiched by two semi-infinite dielectric media of relative permittivities, $\epsilon_{1}$ (orange) and $\epsilon_{2}$ (blue). The graphene sheet is located at the plane defined by $y = 0$ and its electromagnetic properties are encompassed by a conductivity, $\sigma$. Arrows, all lying in a plane called the plane of incidence, which coincides with the $y$-$z$ plane, indicate the propagation directions of the electromagnetic waves (denoted by $A_{1}$ and $A_{2}$ for travelling towards the negative $y$-direction; $B_{1}$ and $B_{2}$, the positive $y$-direction). $\theta_{1}$ and $\theta_{2}$ are the incident and refracted angles, respectively.}
\label{fig:TMM1}
\end{figure}

Let us consider the scattering of electromagnetic waves of frequency, $\omega$, scatter from a planar structure consisting of a monolayer graphene, cladded by two semi-infinite dielectric media of relative permittivities, $\epsilon_{1}$ and $\epsilon_{2}$, as shown in Figure \ref{fig:TMM1}. The graphene sheet is located in the $y = 0$ plane and is assumed to be infinitesimally thin such that medium 1 (characterized by a relative permittivity, $\epsilon_{1}$) occupies the region defined by $y > 0$
and medium 2 (characterized by a relative permittivity, $\epsilon_{2}$) occupies the region defined by $y < 0$. Let us further assume that the waves are plane waves, that their $\mathbf{B}$-field only has an $x$-component, and that their $\mathbf{E}$-field is parallel to the plane of incidence, which coincides with the $y$-$z$ plane (commonly known as transverse magnetic waves or p-polarized waves). Therefore, the $\mathbf{B}$-fields can be written as  

\begin{equation}
\mathbf{B}_{x}^{(j)}(\mathbf{r, t}) = (A_{j}e^{-\mathrm{i}k_{j,y}y} + B_{j}e^{\mathrm{i}k_{j,y}y})e^{\mathrm{i}(k_{j,z}z-\omega t)}\mathbf{\hat{x}},
\label{eq:By_vector}
\end{equation}
where $j = 1, 2$ are for the waves in medium 1 and medium 2, respectively, $A_{j}$ and $B_{j}$ are the amplitudes of the waves propagating in the negative and positive $y$-directions, respectively, subscripts, $x, y,$ and $z$ denote the components associated with the coordinate axes, $t$ is time, $\mathbf{\hat{x}}$ is a unit vector pointing in the positive $x$-direction (out of the page) and the relation between the $y$- and $z$- components of the wavevector is given by 

\begin{equation}
k_{j,y}^{2} = \epsilon_{j}\frac{\omega^{2}}{c^{2}} - k_{j,z}^{2}.
\label{eq:kz_component}
\end{equation}.

In dielectric media, the relation between $\mathbf{B}$- and $\mathbf{E}$- fields is given by the following Maxwell's equation,
\begin{equation}
\nabla \times \mathbf{B} = -\mathrm{i}\frac{\epsilon\omega}{c^{2}}\mathbf{E}.
\label{eq:Maxwell_BE}
\end{equation}

Using the equation given above, the $z$-component of the $\mathbf{E}$-fields is found to be

\begin{equation}
\mathbf{E}_{z}^{(j)}(\mathbf{r, t}) = \frac{-k_{j,y}c^{2}}{\omega \epsilon_{j}}(A_{j}e^{-\mathrm{i}k_{j,y}y} - B_{j}e^{\mathrm{i}k_{j,y}y})e^{\mathrm{i}(k_{j,z}z-\omega t)}\mathbf{\hat{z}},
\label{eq:Ex_vector}
\end{equation}
where $\mathbf{\hat{z}}$ is a unit vector pointing in the positive $z$-direction.\\

In order to find the relations between the amplitudes of the waves in medium 1 and medium 2, we need to invoke the boundary conditions at the interface between the two dielectric media for the parallel components of $\mathbf{E}$- and $\mathbf{B}$- fields, which are

\begin{equation} 
\mathbf{\hat{n}} \times (\mathbf{E}^{(1)} - \mathbf{E}^{(2)}) = 0,
\label{eq:boundary_con_E}
\end{equation}

\begin{equation} 
\mathbf{\hat{n}} \times (\mathbf{B}^{(1)} - \mathbf{B}^{(2)}) = \mu_{0}\mathbf{J}_{s},
\label{eq:boundary_con_B}
\end{equation}
where $\mathbf{\hat{n}}$ is a normal unit vector pointing from medium 2 into medium 1 (equivalent to $\mathbf{\hat{y}}$), and $\mathbf{J}_{s}$ is the free surface current density at the boundary, which is the graphene sheet in this case. Writing out only the relevant components and using the generalized Ohm's law, $\mathbf{J} = \sigma\mathbf{E}$, we obtain

\begin{equation} 
E_{z}^{(1)}(y = 0) = E_{z}^{(2)}(y = 0),
\label{eq:boundary_con_Ex}
\end{equation}

\begin{equation} 
B_{x}^{(1)}(y = 0) - B_{x}^{(2)}(y = 0) = \mu_{0}\sigma E_{z}^{(1)}(y = 0),
\label{eq:boundary_con_By}
\end{equation}
where $\sigma$ is the optical conductivity of the 2D material at the interface (graphene).
Substituting Eqs. \eqref{eq:By_vector} and \eqref{eq:Ex_vector} into the above boundary conditions, yields

\begin{equation} 
A_{1} - B_{1} = \frac{\epsilon_{1}k_{2,y}}{\epsilon_{2}k_{1,y}} (A_{2} - B_{2}),
\label{eq:boundary_A-B}
\end{equation}

\begin{equation} 
A_{1} + B_{1} = (A_{2} + B_{2}) + \frac{\sigma k_{2,y}}{\omega\epsilon_{0}\epsilon_{2}} (A_{2} - B_{2}).
\label{eq:boundary_A+B}
\end{equation}
Converting \eqref{eq:boundary_A-B} and \eqref{eq:boundary_A+B} into a matrix equation yields
\begin{equation} 
\begin{pmatrix}
   1 & -1 \\
   1 & 1
\end{pmatrix}
\begin{pmatrix}
   A_{1} \\
   B_{1}
\end{pmatrix}
=
\begin{pmatrix}
   \eta_{p} & -\eta_{p} \\
   1 + \xi_{p} & 1 - \xi_{p}
\end{pmatrix}
\begin{pmatrix}
   A_{2} \\
   B_{2}
\end{pmatrix},
\label{eq:boundary_matrixform}
\end{equation}
where the following functions have been introduced to simplify our notation,
\begin{equation} 
\eta_{p} = \frac{\epsilon_{1}k_{2,y}}{\epsilon_{2}k_{1,y}} \quad \textrm{and} \quad \xi_{p} = \frac{\sigma k_{2,y}}{\omega\epsilon_{0}\epsilon_{2}}.
\label{eq:eta_xi_p}
\end{equation}

Multiplying \eqref{eq:boundary_matrixform} by the inverse of the leftmost matrix of \eqref{eq:boundary_matrixform}, we obtain a transfer-matrix equation

\begin{align} 
\begin{pmatrix}
   A_{1} \\
   B_{1}
\end{pmatrix} & = \frac{1}{2}
\begin{pmatrix}
   1 + \eta_{p} + \xi_{p} & 1 - \eta_{p} - \xi_{p} \\
   1 - \eta_{p} + \xi_{p} & 1 + \eta_{p} - \xi_{p}
\end{pmatrix}
\begin{pmatrix}
   A_{2} \\
   B_{2}
\end{pmatrix}, \\
& = \mathbf{T}_{p}
\begin{pmatrix}
   A_{2} \\
   B_{2}
\end{pmatrix},
\label{eq:transmission-matrix}
\end{align}
where
\begin{equation} 
\mathbf{T}_{p} = 
\frac{1}{2}
\begin{pmatrix}
   1 + \eta_{p} + \xi_{p} & 1 - \eta_{p} - \xi_{p} \\
   1 - \eta_{p} + \xi_{p} & 1 + \eta_{p} - \xi_{p}
\end{pmatrix}.
\label{eq:transmission-matrix_p}
\end{equation}
Here, $\mathbf{T}_{p}$ is a transmission matrix for p-polarized electromagnetic waves.\\

By following the same procedure, a transmission matrix for s-polarized waves (transverse electric waves) is found to be

\begin{equation} 
\mathbf{T}_{s} = 
\frac{1}{2}
\begin{pmatrix}
   1 + \eta_{s} + \xi_{s} & 1 - \eta_{s} + \xi_{s} \\
   1 - \eta_{s} - \xi_{s} & 1 + \eta_{s} - \xi_{s}
\end{pmatrix},
\label{eq:tranmission-matrix_s}
\end{equation}
where
\begin{equation} 
\eta_{s} = \frac{k_{2,y}}{k_{1,y}} \quad \textrm{and} \quad 
\xi_{s} = \frac{\sigma \mu_{0} \omega}{k_{1,y}}.
\label{eq:eta_xi_s}
\end{equation}

The propagation matrix can easily be derived by recalling the fact that an electromagnetic wave with $y$-component of wavevector, $k_{y}$, propagating through a distance, $d$, in the $y$-direction in a uniform medium only changes its phase by $k_{y}d$. Hence, the propagation matrix is given by
\begin{equation} 
\mathbf{P}(d) = 
\begin{pmatrix}
   e^{-ik_{y}d} & 0 \\
   0 & e^{ik_{y}d}
\end{pmatrix}.
\label{eq:propagation-matrix}
\end{equation}

Now that we have obtained explicit forms for both transmission and propagation matrices, a transfer-matrix for calculating the reflection coefficients can be constructed from a series of matrix multiplications in reverse chronological order of the scattering events that the matrices correspond to.\\

Writing the transfer-matrix in the form $\mathbf{M} = \begin{pmatrix}
   M_{11} & M_{12} \\
   M_{21} & M_{22}
\end{pmatrix}$, the reflection coefficients can be straightforwardly obtained from two of the matrix elements via
\begin{equation} 
r_{s,p} = \frac{M_{21}}{M_{11}}.
\label{eq:reflection_coeff}
\end{equation}

For the atom-chip structure shown in Fig.~\ref{fig:chip_diagram}, the associated transfer-matrix can be written as

\begin{equation} 
\mathbf{M} = \mathbf{T}_{\mathrm{hBN}}\mathbf{P}(t_{\mathrm{hBN}})\mathbf{T}_{\mathrm{Gr}}\mathbf{P}(t_{\mathrm{hBN}})\mathbf{T}_{\mathrm{hBN}},
\label{eq:transfer-matrix}
\end{equation}
where $\mathbf{T}_{\mathrm{hBN}}$ and $\mathbf{T}_{\mathrm{Gr}}$ are, respectively, associated with transmission across the hBN interface and the graphene interface, while $\mathbf{P}(t_{\mathrm{hBN}})$ corresponds to propagation through the thickness, $t_{\mathrm{hBN}}$, of the hBN layer.

\section{Three-body loss rate in 1D BECs}
\label{sec:3b_loss_appendix}
In this section, we provide a more detailed derivation of the three-body loss rate considered in the main text, which mainly follows \cite{schemmer2018,bouchoule_schemmer_henkel_2018, pethick_smith_2008, petrov_gangardt_2004}. 
We start by considering an \textsuperscript{87}Rb quasi-condensate with a mean atomic volume density, $\rho_{0}(\mathbf{r}, t)$, at time $t$, where $\mathbf{r} = (x,y,z)$ denotes the center of a cell of volume, $\Delta$, which is small enough for the condensate to be considered homogeneous throughout the cell (i.e. $\partial\rho_{0}(\mathbf{r}, t)/\partial \Delta \approx 0$), but also large enough to accommodate many atoms.
Assuming that the condensate is subject to a three-body loss process, its mean volume density evolves in time according to \cite{bouchoule_schemmer_henkel_2018,schemmer2018}

\begin{equation} 
\frac{\mathrm{d}\rho_{0}}{\mathrm{d}t} = -\kappa_{\mathrm{Rb}}\rho_{0}^{3},
\label{eq:rate of change of mean volume density}
\end{equation} 
where $\kappa_{\mathrm{Rb}} = \SI{1.8e-41}{m^{6}s^{-1}}$ is the three-body recombination rate for \textsuperscript{87}Rb atoms in the $F = m_{F} = 2$ state \cite{soding1999}, and we have dropped the explicit dependence of $\rho_{0}$ on $(\mathbf{r}, t)$ for simplicity. Henceforth, we derive the three-body loss rate at the center of the trap ($z = 0$) for a 1D quasi-condensate by integrating the loss rate for a 3D quasi-condensate given in Eq. \eqref{eq:rate of change of mean volume density} over the radial coordinate.\\

But first, let us assume that this condensate is greatly elongated in one dimension, i.e. that it is trapped in a smoothly-varying anisotropic harmonic potential with radial trapping frequency $\omega_{r} = \omega_{x} = \omega_{y}$ and axial trapping frequency $\omega_{z}$, where $\omega_{r} >> \omega_{z}$. Consequently, the density profile of the condensate can be described by a one-dimensional Thomas-Fermi distribution in the $z$-direction, multiplied by the Gaussian ground-state quantum harmonic oscillator wavefunctions in the $x$- and $y$- directions \cite{pethick_smith_2008}: 

\begin{equation} 
\rho_{0}(r, z) = \frac{1}{U_{0}}\Big(\mu_{\mathrm{eff}} - \frac{m\omega_{z}^{2}}{2}z^{2}\Big)e^{-r^{2}/2a_{r}^2},
\label{eq:Thomas-Fermi distribution}
\end{equation} 
where $U_{0} = 4\pi\hbar^{2}a_{T}/m$, $a_{T} = \SI{5.6}{\nano\metre}$ is the s-wave scattering length \cite{roberts_1998}, $\mu_{\mathrm{eff}} = \mu - \hbar\omega_{r}$, $\mu$ is the chemical potential of the condensate, $m = \SI{1.44e-25}{kg}$ is the mass of an \textsuperscript{87}Rb atom, $a_{r} = \sqrt{\hbar/m\omega_{r}}$, and $r = \sqrt{x^{2} + (y-y_{c})^{2}}$ is the radial distance measured from the mean positions of the harmonic oscillator states at $x = 0$ and $y = y_{c}$. It can be seen that in the $z$-direction, the mean volume density peaks at the trap center, where $z = 0$.\\ 


We can obtain the mean line density, $n_{0}(z)$, by integrating Eq. \eqref{eq:Thomas-Fermi distribution} radially from $r = 0$ to $r = \infty$:

\begin{equation} \label{eq:line_density}
\begin{split}
\int_{0}^{\infty}\rho_{0}2\pi r \mathrm{d}r &= \frac{1}{U_{0}}\Big(\mu_{\mathrm{eff}} - \frac{m\omega_{z}^{2}}{2}z^{2}\Big)\int_{0}^{\infty}e^{-r^{2}/2a_{r}^2}2\pi r \mathrm{d}r,\\
n_{0}(z) &= \frac{1}{U_{0}}\Big(\mu_{\mathrm{eff}} - \frac{m\omega_{z}^{2}}{2}z^{2}\Big)\Big[0 + 2\pi a_{r}^{2}\Big],\\
n_{0}(z) &= \frac{2\pi a_{r}^{2}}{U_{0}}\Big(\mu_{\mathrm{eff}} - \frac{m\omega_{z}^{2}}{2}z^{2}\Big).
 \end{split}
\end{equation}

Performing a radial integration on Eq. \eqref{eq:rate of change of mean volume density} in the same manner, we obtain the time evolution of the mean line density as follows

\begin{equation} \label{eq:1D loss rate}
\begin{split}
\int_{0}^{\infty}\frac{\mathrm{d}\rho_{0}}{\mathrm{d}t}2\pi r \mathrm{d}r &= -\int_{0}^{\infty}\kappa_{\mathrm{Rb}}\rho_{0}^{3}2\pi r \mathrm{d}r,\\
\frac{\mathrm{d}n_{0}(z)}{\mathrm{d}t} &= -\frac{\kappa_{\mathrm{Rb}}}{12 \pi^{2} a_{r}^{4}}n_{0}(z)^{3},
\end{split}
\end{equation}
where we have used Eqs. \eqref{eq:Thomas-Fermi distribution} and \eqref{eq:line_density} to convert from a volume to a line atomic density.\\
\\

\bibliographystyle{unsrt}
\bibliographystyle{apsrev4-2}
\bibliography{NewRef}


\end{document}